\documentclass[11pt]{article}

\usepackage[T1]{fontenc}
\usepackage{lmodern}
\usepackage[margin=1in]{geometry}
\usepackage{amsmath,amssymb,amsthm}
\usepackage{subcaption}
\usepackage[hidelinks]{hyperref}
\usepackage{tikz}
\usetikzlibrary{arrows.meta}
\definecolor{overviewink}{HTML}{1C2735}
\definecolor{overviewslate}{HTML}{566577}
\definecolor{overviewblue}{HTML}{17648A}
\definecolor{overvieworange}{HTML}{9B5421}
\definecolor{overviewfill}{HTML}{EFF6FA}

\newtheorem{theorem}{Theorem}
\newtheorem{lemma}[theorem]{Lemma}
\newtheorem{proposition}[theorem]{Proposition}
\theoremstyle{definition}
\newtheorem{corollary}[theorem]{Corollary}
\newtheorem{definition}[theorem]{Definition}
\newtheorem*{definition*}{Definition}
\newtheorem*{remark}{Remark}

\newcounter{procedure}
\newenvironment{pseudocode}{\par\addvspace{\medskipamount}\noindent
  \begin{minipage}{\linewidth}\small
  \hrule\smallskip
}{\smallskip\hrule
  \end{minipage}\par\addvspace{\medskipamount}}
\newcommand{\pseudocodetitle}[2]{\refstepcounter{procedure}\label{#1}\textbf{Algorithm~\theprocedure.} #2\par
}

\newcommand{\E}{\mathbb{E}}
\newcommand{\OPT}{\mathrm{OPT}}
\newcommand{\conv}{\operatorname{conv}}

\title{Optimal Online Hitting Set for Disks via Delaunay Graphs}
\author{Mingchao Zhou\thanks{School of Mathematical Sciences, Zhejiang Normal University, Jinhua, China.  \texttt{zmc138575@gmail.com, hxhzz@sina.com}.}
  \and
  Yixin Cao\thanks{Department of Computing, Hong Kong Polytechnic University, Hong Kong, China. \texttt{yixin.cao@polyu.edu.hk}. }
  \and
  Zhao Zhang\footnotemark[1]
}
\date{}

\begin{document}
\maketitle

\begin{abstract}
We give a randomized algorithm with an $O(\log n)$ expected competitive ratio for online hitting set with disks of arbitrary radii.  It irrevocably selects $O(k(1+\log(n/k)))$ points from $n$ planar candidates in expectation on every fixed nonempty request sequence with offline optimum~$k$.  The disk Delaunay graph is planar on every candidate subset~$S$ and therefore has at most $3|S|$ edges.  More generally, if the Delaunay graph on each subset~$S$ has at most $\kappa|S|$ edges, the algorithm selects $O((1+\kappa)k(1+\log(n/k)))$ points in expectation.  The framework also applies to shortest-path balls in graphs that exclude a fixed minor.  A matching lower bound for every $k$ holds for online interval hitting and therefore also for disks on collinear candidates and shortest-path balls on paths.  For disks given by center and radius, inversion and convex hulls yield $O(n^2\log n)$ geometric preprocessing operations, and a balanced tree with Voronoi diagrams supports $O(\log^2 n)$ worst-case time per request.
\end{abstract}

\section{Introduction}

We study online hitting set for disks of arbitrary radii in the plane.  The algorithm receives a set~$P$ of $n$ candidate points in advance.  Each request is a disk~$D$ that meets $P$, specified by its center and radius.  After every arrival, the algorithm must maintain a hitting set~$X\subseteq P$ for all disks seen so far, and it can only add points to $X$.  We compare the final size of $X$ with the offline optimum~$k$, the size of a smallest hitting set for the entire sequence.

With $n$ candidates and $m$ distinct subsets that can arrive as requests, the dual set-cover instance has $n$ sets and $m$ elements.  The online set-cover algorithm in~\cite{alon} gives an $O(\log n\log m)$ competitive ratio.  Disks have bounded VC dimension, so they induce only polynomially many distinct subsets of~$P$, giving an $O(\log^2 n)$ ratio.  For general online hitting set, the result of Doron-Arad and Naor~\cite{doron} implies an $\Omega(\sqrt n)$ lower bound on the expected competitive ratio against an oblivious adversary, even for randomized algorithms that know the full family of possible requests in advance.

Even and Smorodinsky~\cite{even} obtained an $O(\log n)$ competitive ratio for unit disks and asked whether one could improve the $O(\log^2 n)$ bound for arbitrary radii.  We answer this question affirmatively.

\begin{theorem}\label{thm:disk}
Let $P$ be a set of $n$ points in the plane.  There is a randomized online algorithm for hitting closed disks of arbitrary radii that meet $P$.  On every fixed nonempty request sequence, its expected cost is $O(k(1+\log(n/k)))$, where $k$ is the offline optimum.
\end{theorem}

The dependence on the offline optimum~$k$ matters when it is large.  For example, when $k=\Theta(n/\log n)$, the bound~$O(k\log n)$ is already $O(n)$, the cost of selecting all points.  Our algorithm selects $O(k(1+\log(n/k)))$ points in expectation, giving $O(k\log\log n)$ in this case.

For disks specified by center and radius, the algorithm uses $O(n^2\log n)$ geometric operations in preprocessing and $O(\log^2 n)$ worst-case time per request.

Even and Smorodinsky~\cite[Proposition~6 and Theorem~22]{even} used nested interval requests to prove a logarithmic lower bound for deterministic algorithms with offline optimum one and realized those requests with unit disks.  We repeatedly choose either half of an interval with equal probability and realize the resulting nested requests as disks.  This gives the following matching lower bound for randomized algorithms.

\begin{theorem}\label{thm:disk-lower}
For all integers~$n,k$ with $1\le k\le n$, there is a set~$P$ of $n$ collinear points such that, for every randomized online algorithm on $P$, some fixed sequence of disk requests with offline optimum~$k$ forces the algorithm to select at least $k(1+\tfrac12\lfloor\log_2(n/k)\rfloor)$ points in expectation.
\end{theorem}

On collinear candidates, disk hitting reduces to interval hitting.  Appendix~\ref{app:refined-intervals} gives the exact randomized worst-case expected cost when $k=1$ and an algorithm for arbitrary interval sequences with expected competitive ratio $\tfrac12\log_2 n+O(\sqrt{\log n})$.  A single algorithm achieves this ratio for every $k$, choosing its parameter using only $n$.  The leading constant $1/2$ matches the lower bound.

For each subset~$S\subseteq P$, join two points by an edge when some disk meets $S$ in exactly those two points.  This Delaunay graph is planar for every $S$, so it has at most $3|S|$ edges.  Our algorithm uses a bound of $\kappa|S|$ on the number of edges in each such graph and achieves expected cost~$O((1+\kappa)k(1+\log(n/k)))$.  For disks, hereditary planarity gives $\kappa=3$.  The same bound extends to shortest-path balls in graphs that exclude a fixed minor.

\begin{theorem}\label{thm:graph}
Fix a graph~$F$.  Let $H$ be a finite connected undirected graph that excludes $F$ as a minor and has positive rational edge lengths.  Let $P\subseteq V(H)$ be a set of $n$ candidate vertices.  There is a randomized online algorithm for hitting shortest-path balls that meet $P$.  On every fixed request sequence with offline optimum~$k\ge 1$, the algorithm selects $O(k(1+\log(n/k)))$ vertices in expectation, with a constant depending only on $F$.
\end{theorem}

For each $1\le k\le n$, a matching lower bound already holds on paths, so the cost bound is tight for planar graphs (Corollary~\ref{cor:path-lower}).

\subsection{Related work}\label{sec:related}
Even and Smorodinsky~\cite{even} used unique-maximum colorings to obtain tight logarithmic competitive ratios for online hitting of intervals, halfplanes, and unit disks.
De, Singh, and T\'oth~\cite{deesa} extended the unit-disk setting to radii in $[1,M]$, giving a deterministic $O((1+\log M)\log n)$ competitive ratio that matches the unit-disk bound at $M=1$.  Their bounded-scale result also applies to positive homothets of any fixed planar convex body.  They obtained an $O(\log n)$ competitive ratio for disks of arbitrary radii when a given line separates the candidate points from the disk centers~\cite[Theorem~9]{deesa}.  The same authors~\cite{deswat} gave a deterministic $O(\log n)$-competitive algorithm for axis-aligned squares of arbitrary size on any set of $n$ points in the plane. They extended this result to positive homothets of a fixed polygon with $t\ge3$ vertices, obtaining a deterministic competitive ratio of $O(t^2\log n)$.

De, Mandal, and Singh~\cite[Theorem~21]{dms-polygons} gave a randomized $O(t^2(1+\log M)\log n)$-competitive algorithm for homothetic copies of a regular $t$-gon, $t\ge4$, with diameters in $[1,M]$.  Alefkhani, Khodaveisi, and Mari~\cite{akm} gave a deterministic $O(\log N)$-competitive algorithm for disks contained in $(0,N)^2$ when the candidate set is $P=(0,N)^2\cap\mathbb Z^2$.  For axis-parallel squares, the authors of~\cite{khan} obtained a deterministic $O(\log N)$ competitive ratio when all candidate points and square vertices have integer coordinates in $[0,N)^2$.  For a known family of $m$ possible ranges, each containing at most $d$ candidates, the online set-cover algorithm of Buchbinder and Naor~\cite[Section~5.1]{bn} yields, by duality, a deterministic online hitting-set algorithm that selects $O(k(1+\log d)(1+\log(m/k)))$ points.

For the related problem of maintaining an $\epsilon$-net online, Bhore, Dey, and Singh~\cite[Theorem~2]{bhore-net} obtained a deterministic $O(1+\log(1/\epsilon))$ competitive ratio for intervals.  They also gave a randomized construction for online $\epsilon$-nets for planar axis-parallel rectangles \cite[Theorem~3]{bhore-net}.  Bhore, Gupta, and Kumar~\cite{bhore} proposed using $\epsilon$-nets to round a fractional online hitting set after replacing each point with copies according to its fractional value.  The paper was retracted after the authors found that the shallow-cell complexity of the resulting set system could be much larger than that of the original, invalidating the claimed competitive bounds.

The set-cover reduction gives an $O(\log^2 n)$ competitive ratio directly from bounded VC dimension of the original range family.  A family of bounded VC dimension can still have a dense Delaunay graph.  For example, on $n\ge 4$ candidates the range family of all pairs has VC dimension two and its Delaunay graph on $P$ is complete.

\subsection{Technical overview}\label{sec:overview}

For $n\ge 2$, the algorithm samples orders during randomized preprocessing.  Given those orders, online selection is deterministic.
\begin{description}
\item[{Offline preprocessing.}] Before any requests arrive, use multiplicative weights to sample an order~$\pi_s$ on $P$ for each cost guess~$s=1,2,4,\ldots$ with $s<n$.  Set the disk phase budget $B_s=72s(1+\lceil\log_2(n/s)\rceil)$.
\item[{Online selection.}] Start with $s=1$.  For each arriving disk, find its first point in $\pi_s$.
  If the point would be the $(B_s+1)$st distinct first point in the current phase, start a new phase with guess $2s$ and find the first point of the same disk in the new order.
  Select the resulting point.
\end{description}

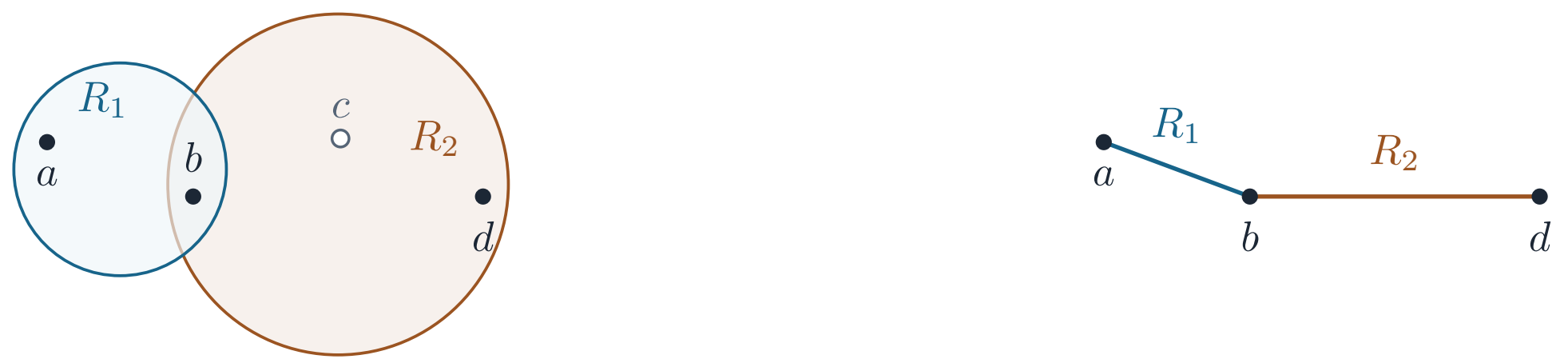
\begin{figure}[!htbp]
  \centering
  \begin{subfigure}[t]{.47\linewidth}
    \centering
    \begin{tikzpicture}[font=\small,text=overviewink]
      \useasboundingbox (-2.35,-1.42) rectangle (2.35,1.62);
      \draw[overvieworange,line width=.7pt,fill=overvieworange!8]
        (.6,.1) circle[radius=1.41];
      \draw[overviewblue,line width=.7pt,fill=overviewfill,
        fill opacity=.65] (-1.205,.225) circle[radius=.88];
      \node[text=overviewblue] at (-1.35,.8) {$R_1$};
      \node[text=overvieworange] at (1.4,.48) {$R_2$};
      \foreach \x/\y/\point in {-1.81/.45/a,1.8/0/d} {
        \fill[overviewink] (\x,\y) circle[radius=1.9pt]
          node[below=2pt] {$\point$};
      }
      \fill[overviewink] (-.6,0) circle[radius=1.9pt] node[above=2pt] {$b$};
      \draw[overviewslate,fill=white,line width=.65pt]
        (.62,.48) circle[radius=2pt] node[above=1pt,text=overviewslate] {$c$};
    \end{tikzpicture}
    \caption{On $S=\{a,b,d\}$, the traces are $\{a,b\}$ and $\{b,d\}$.}
  \end{subfigure}\hfill
  \begin{subfigure}[t]{.47\linewidth}
    \centering
    \begin{tikzpicture}[font=\small,text=overviewink]
      \useasboundingbox (-2.35,-1.42) rectangle (2.35,1.62);
      \draw[overviewblue,line width=1pt] (-1.81,.45) -- node[above=2pt,text=overviewblue] {$R_1$} (-.6,0);
      \draw[overvieworange,line width=1pt] (-.6,0) -- node[above=2pt,text=overvieworange] {$R_2$} (1.8,0);
      \foreach \x/\y/\point in {-1.81/.45/a,-.6/0/b,1.8/0/d} {
        \fill[overviewink] (\x,\y) circle[radius=1.9pt]
          node[below=2pt] {$\point$};
      }
    \end{tikzpicture}
    \caption{The edges of $G(S)$ are $ab$ and $bd$.}
  \end{subfigure}
  \caption{The ranges on $P=\{a,b,c,d\}$ are $R_1=\{a,b\}$ and $R_2=\{b,c,d\}$.  In the order~$\langle a,b,c,d\rangle$, their first points are $a$ and $b$, respectively.  The hollow point~$c$ is not in $S$.}
  \label{fig:definition-traces}
\end{figure}

First consider a fixed order~$\pi$ on $P$.  If $\pi$ starts with an optimal hitting set for a fixed request sequence, the distinct first points are exactly the points of that hitting set.  The design question is how to sample an order with small expected cost before the requests arrive.  In Figure~\ref{fig:definition-traces}, for example, the order~$\langle a,b,c,d\rangle$ selects $a$ from $R_1$ and $b$ from $R_2$, while $\{b\}$ hits both ranges.

To bound the cost of a sampled order, fix an offline hitting set~$C$.  For each point~$p\in C$, count how many distinct candidates can be first in a disk containing~$p$.  Every selected point is counted because a request for which it is first contains a member of~$C$.  A first point may be counted for several members of~$C$, so the sum bounds the number of distinct selected points.

If $p$ is the first candidate of a disk~$D$ in $\pi$, then for every other candidate~$q\in D\cap P$, the disk meets $S=B_\pi(p)\cup\{q\}$ in exactly $\{p,q\}$.  Here $B_\pi(p)$ is the set of points in the prefix of $\pi$ ending at $p$.  The pair~$\{p,q\}$ is a \emph{two-point trace} and gives an edge of the Delaunay graph~$G(S)$.  In Figure~\ref{fig:definition-traces}, $b$ is first in $R_2$, which witnesses the edge~$bd$ on $S=B_\pi(b)\cup\{d\}=\{a,b,d\}$.  When counting possible first points for a fixed $q$, each candidate~$p$ has its own prefix~$B_\pi(p)$.  The analysis therefore needs sparsity on every restriction.  For disks, $G(S)$ is planar and has at most $3|S|$ edges.

\begin{figure}[!htbp]
\centering
  \begin{subfigure}[t]{.51\linewidth}
\centering
    \begin{tikzpicture}[x=.65cm,font=\small,text=overviewink]
      \draw[-{Latex[length=2mm]},line width=.65pt,overviewslate] (0,0) -- (9.5,0);
      \draw[overviewblue!65,dashed,line width=.6pt] (5.55,0) -- (5.55,-.3)
        node[below,text=overviewblue] {$t$};
      \foreach \position/\point in {1.2/a,3.25/b,7.1/c} {
        \fill[overviewslate] (\position,0) circle[radius=1.8pt] node[above=2pt] {$\point$};
      }
      \node[fill=overviewblue,minimum size=4.6pt,inner sep=0pt,label={[text=overviewblue]above:$p$}] at (5.55,0) {};
      \node[circle,draw=overvieworange,line width=.85pt,fill=white,minimum size=5.4pt,inner sep=0pt,label={[text=overvieworange]above:$q$}] at (8.45,0) {};
    \end{tikzpicture}
    \caption{When only $p$ rings at $t<\tau_q$, the test uses $S_t\cup\{q\}$.}
    \label{fig:overview-clocks-a}
  \end{subfigure}\hfill
  \begin{subfigure}[t]{.47\linewidth}
\centering
    \begin{tikzpicture}[x=.65cm,font=\small,text=overviewink]
      \draw[-{Latex[length=2mm]},line width=.65pt,overviewslate] (0,0) -- (8.5,0);
      \draw[overviewblue!65,dashed,line width=.6pt] (5.55,0) -- (5.55,-.3)
        node[below,text=overviewblue] {$t$};
      \foreach \position/\point in {1.2/a,3.25/b,7.1/c} {
        \fill[overviewslate] (\position,0) circle[radius=1.8pt] node[above=2pt] {$\point$};
      }
      \node[fill=overviewblue,minimum size=4.6pt,inner sep=0pt,label={[text=overviewblue]above:$p$}] at (2.15,0) {};
      \node[circle,draw=overvieworange,line width=.85pt,fill=white,minimum size=5.4pt,inner sep=0pt,label={[text=overvieworange]above:$q$}] at (4.6,0) {};
    \end{tikzpicture}
    \caption{With $\tau_p,\tau_q\le t$, the test uses $S_t$.}
    \label{fig:overview-clocks-b}
  \end{subfigure}
  \caption{The clocks of $a,b,c$ agree in both panels.  The test for the edge~$pq$ uses the same vertex set~$\{a,b,p,q\}$ in both cases.}
  \label{fig:overview-clocks}
\end{figure}
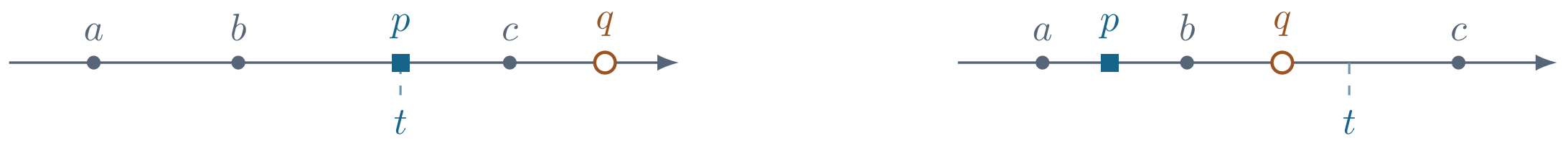

To compare these graphs, we use exponential clocks~\cite{bns}.  Give each point~$p$ an independent clock that rings once after an exponentially distributed waiting time~$\tau_p$.  For $t\ge 0$, let $S_t$ be the points whose clocks have rung by time~$t$.  With probability one the ringing times are distinct.  On this event, when $p$ rings at $t$ before $q$, the edge test is on $S_t\cup\{q\}=S_t\cup\{p,q\}$, as in Figure~\ref{fig:overview-clocks-a}.  Including both endpoints makes the test depend only on the other clocks.  When both endpoints have rung by $t$, the same test asks whether $pq$ is an edge of $G(S_t)$, as in Figure~\ref{fig:overview-clocks-b}.  This allows us to bound the weighted contribution at each time in terms of the expected edge count in a single sparse graph.

For each guess~$s$, multiplicative weights turn the weighted load bound into an expected cost guarantee for every fixed sequence that can be hit by at most $s$ points.  The analysis bounds the total cost across phases using their budgets and the probabilities of reaching them.

For a fixed order on $P$, permuting the requests preserves the set of selected points.  In the phased algorithm, the arrival order can change which requests are processed under each order.  A disk that arrives again in a later phase can yield a different first point.  See the arrival-order example at the end of Section~\ref{sec:general}.

The charging argument uses only the points in each request and the pairs it isolates on candidate subsets.  For disks, the graph of these pairs is planar.  For shortest-path balls, contracting the Voronoi cells of a nonempty candidate subset~$S$ gives a minor of the host graph containing $G(S)$.  Excluding a fixed minor therefore gives a linear edge bound on every subset, just as for disks.  This lets us apply the same algorithm and analysis.

For disk requests, geometric inversion reduces each sampling round's batch of Delaunay edge tests to membership queries in one convex hull (Section~\ref{sec:disk-implementation}).  During online selection, a balanced tree with Voronoi diagrams at its nodes finds the first point of $D\cap P$ in the current order without enumerating the range.  A phase change can require a second search, under the new order (Section~\ref{sec:disk-first-point}).

\section{Online hitting set with sparse Delaunay graphs}\label{sec:general}

This section presents our framework, which uses an oracle that tests adjacency in $G(S)$.  Its implementation will be furnished in Sections~\ref{sec:disks} and~\ref{sec:graph}.

\subsection{Range spaces, two-point traces, and Delaunay graphs}\label{sec:prelim}

Let $(P,\mathcal R)$ be a range space, where $P$ is the set of $n$ candidate points and $\mathcal R\subseteq 2^P\setminus\{\varnothing\}$ is the family of possible \emph{ranges}.  A disk or ball~$D$ that meets $P$ induces the range~$D\cap P$.  A disk request is specified by its center and radius, from which membership must be determined.  In the abstract formulation, each request is an explicit range given as a list of its points.  A range can occur more than once, since even distinct disks can induce the same subset of $P$.

A \emph{request sequence} is a finite ordered list~$R_1,\ldots,R_\ell$ of ranges in $\mathcal R$, with duplicates allowed.  The algorithm knows $P$ in advance and receives the requests one at a time in that order.  It starts with $X=\varnothing$ and can only add points from $P$ to $X$.  After processing request~$R_i$, it must have $X\cap R_j\ne\varnothing$ for every $1\le j\le i$.  Its cost is the final size~$|X|$.  Let $\OPT$ be a minimum hitting set for the entire sequence, and write $k=|\OPT|$ for the offline optimum.

A randomized online algorithm is \emph{$f(n)$-competitive} against an oblivious adversary if, for every nonempty request sequence fixed before the algorithm runs, it satisfies $\E[|X|]\le f(n)k$.  The expectation is over the algorithm's random choices.

An empty sequence costs nothing.  For $n=1$ it is optimal to select the sole point on the first request.  The algorithm below assumes $n\ge 2$.

\begin{samepage}
\begin{definition}[Two-point traces and the Delaunay graph]
For $S\subseteq P$, the \emph{trace} of a range~$R$ on $S$ is $R\cap S$.  Restricting $(P,\mathcal R)$ to $S$ gives a range space on $S$ whose ranges are the distinct nonempty traces of members of $\mathcal R$.  A \emph{two-point trace} on $S$ is a trace~$R\cap S$ of cardinality two.  The \emph{Delaunay graph}~$G(S)$ has vertex set~$S$ and one edge for each two-point trace on $S$.\label{def:two-point-traces}
\end{definition}
\end{samepage}

Our definition follows Ackerman, Keszegh, and P\'alv\"olgyi~\cite[Definition~2]{ackerman}.  The name refers to Delaunay's empty-sphere construction~\cite{delaunay1934}.  Figure~\ref{fig:definition-traces} illustrates the definition.  The range~$R_2$ has two-point trace~$\{b,d\}$ on $S$ even though it also contains $c\notin S$.

For brevity, we often write $p q\in G$ instead of $p q\in E(G)$.

\subsection{Orders, the blind rule, and load}

Fix an order~$\pi$ on $P$.  Let $B_\pi(p)$ be the set of points in the prefix of $\pi$ ending at $p$.  For $q$ at or after $p$ in $\pi$, define
\[
  G_\pi(p,q)=G\bigl(B_\pi(p)\cup\{q\}\bigr).
\]
If $p$ precedes $q$ in $\pi$, then $pq\in G_\pi(p,q)$ exactly when some range in $\mathcal R$ has trace~$\{p,q\}$ on $B_\pi(p)\cup\{q\}$.  Such a range contains both points and misses every point before $p$, making $p$ its first point.  Conversely, any range containing $q$ with first point~$p$ has that trace.

For the convenience of analysis, we use the \emph{blind rule}, which adds the first point of each arriving range~$R$ in $\pi$ to the hitting set, even if a previously selected point already hits $R$.  With a fixed $\pi$, the final hitting set consists of the distinct first points of the requests, so it is independent of their arrival order.
With $\pi=\langle a,b,c,d\rangle$, the blind rule selects $\{a,b\}$ on the two requests in Figure~\ref{fig:definition-traces} in either arrival order.  This reduces the cost analysis to counting distinct first points.

For a nonempty request sequence, let $q$ be the first point of $\pi$ for which $H=B_\pi(q)$ hits every request.  The blind rule selects only points in $H$, including $q$ because some request misses $H\setminus\{q\}$.  If $H$ is inclusion-minimal, each $p\in H$ is the only point of $H$ in some request.  Since $H$ is a prefix of~$\pi$, $p$ is the first point of that request.  Thus the blind rule selects all of $H$.  In particular, if $\pi$ starts with the points of $\OPT$, then $H=\OPT$ because no shorter prefix can hit every request, and the selected set is exactly $\OPT$ at cost~$k$.

Fix a prefix~$A$ of $\pi$, consisting of the points in the first $|A|$ positions, and a comparison hitting set~$C$ of the request sequence.  For a selected point~$p\not\in A$, any request whose first point is $p$ misses $A$.  We charge the cost of $p$ to a point in $C\setminus A$.  The algorithm does not select $A$ automatically.  We count the selected points in $A$ separately; there are at most $|A|$ such points.  The following definition bounds how many distinct selections can be charged to each point.
\begin{definition}[Load]
For $p\in A$, set $L_{\pi,A}(p)=0$.  For $p\notin A$, the \emph{load}~$L_{\pi,A}(p)$ reserves one unit for the possible selection of $p$ itself and counts each earlier point~$q\notin A$ that can be first in a range containing $p$.  Thus
\begin{equation}\label{eq:load}
  L_{\pi,A}(p)
  =1+\bigl|\{q\in B_\pi(p)\setminus A: pq\in G_\pi(q,p)\}\bigr|.
\end{equation}
The graph~$G_\pi(p,p)=G(B_\pi(p))$ has no self-loops, so the set in \eqref{eq:load} counts only earlier points.
\end{definition}

\begin{samepage}
In Figure~\ref{fig:definition-traces}, take $\pi=\langle a,b,c,d\rangle$.  The first point of $R_1$ is $a$, and the first point of $R_2$ is $b$.
\begin{center}
  \begin{tabular}{@{}clcc@{}}
    Point $p$ & Earlier first points & $L_{\pi,\varnothing}(p)$ & $L_{\pi,\{a\}}(p)$ \\
    \hline
    $a$ & None & $1$ & $0$ \\
    $b$ & $a$ via $R_1$ & $2$ & $1$ \\
    $c$ & $b$ via $R_2$ & $2$ & $2$ \\
    $d$ & $b$ via $R_2$ & $2$ & $2$
  \end{tabular}
\end{center}
With $A=\{a\}$, the load at $a$ is zero and $a$ is excluded from the load on $b$.  The unit term applies to every point outside $A$, so $c$ and $d$ each have load~$1+1=2$ in both cases.
\end{samepage}

\begin{lemma}\label{lem:charge}
Fix an order~$\pi$ and a prefix~$A$ of $\pi$.  For any hitting set~$C$ of the request sequence, the blind rule selects a set of size at most
\[
  |A|+\sum_{p\in C} L_{\pi,A}(p).
\]
\end{lemma}
\begin{proof}
  Let $S$ be the set of selected points.
  For each point~$p\in S\setminus A$, choose a request~$R$ whose first point in $\pi$ is $p$.  Since $A$ consists of points before $p$, the range~$R$ misses $A$.  If $p\in C$, charge $p$ to itself.  The unit term in \eqref{eq:load} accounts for this charge.  Otherwise, choose any $q\in C\cap R$ and charge $p$ to $q$.  Since $p$ is first in $R$, it precedes $q$ in $\pi$ and $R$ witnesses $pq\in G_\pi(p,q)$.  Because $q\notin A$ and~$p$ is an earlier point outside $A$, the load on $q$ counts $p$.  Each distinct selected point outside $A$ is charged once, and at most $|A|$ selected points lie in $A$.
\end{proof}

\subsection{Sampling an order for a fixed guess}

Fix a guess~$s\in\{1,\ldots,n\}$ for the offline optimum~$k$.  We seek an order with small expected total load on any fixed set of at most $s$ points.  In each round, \textsc{WeightedOrder} takes positive integer weights~$u(p)$ and samples an order~$\pi$ and a prefix~$A$.  For $X\subseteq P$, write $u(X)=\sum_{p\in X}u(p)$.  We bound the expected size of $A$ and the expected weighted load
\[
  \mu\cdot L_{\pi,A}=\sum_{p\in P}\mu_p L_{\pi,A}(p), \text{ where } \mu_p= \frac{u(p)}{u(P)}.
\]
The outer procedure, \textsc{SampleOrder}, updates the weights between rounds to control the total load on a hitting set.

\begin{pseudocode}
  \begin{minipage}[t]{.48\linewidth}
    \hrule width\linewidth height0pt\relax
    \begin{tabbing}
      0.\quad\=1.1.\quad\=\quad\=\kill
      0.\> $\pi\gets$ the empty sequence\\
      1.\> \textbf{for} $p\in P$ in a fixed order \textbf{do}\\
      \> 1.1.\> Append $p$ to $\pi$ independently\\
      \>\> with probability $\min\{1,su(p)/u(P)\}$\\
      2.\> $A\gets$ points of $\pi$, $M\gets P\setminus A$\\
      3.\> \textbf{while} $M\ne\varnothing$ \textbf{do}\\
      \> 3.1.\> Draw $q\in M$ with prob.\ $u(q)/u(M)$\\
      \> 3.2.\> Append $q$ to $\pi$ and remove it from $M$\\
      4.\> \textbf{return} $(\pi,A)$
    \end{tabbing}
  \end{minipage}\hfill
  \begin{minipage}[t]{.48\linewidth}
    \hrule width\linewidth height0pt\relax
    \begin{tabbing}
      0.\quad\=1.1.\quad\=\quad\=\kill
      1.\> \textbf{for} $p\in P$ \textbf{do} $u_1(p)\gets 1$\\
      2.\> \textbf{for} $j=1,\ldots,2n$ \textbf{do}\\
      \> 2.1.\> $(\pi_j,A_j)\gets \text{\textsc{WeightedOrder}}(s,u_j)$\\
      \> 2.2.\> Draw $Q$ uniformly from $P$\\
      \> 2.3.\> \textbf{for} $p\in P$ \textbf{do}\\
      \>\>\> Compute $\widehat\ell_j(p)$ by \eqref{eq:estimate}\\
      \>\>\> $u_{j+1}(p)\gets(n+1)(1+\widehat\ell_j(p))u_j(p)$\\
      3.\> Draw $J$ uniformly from $\{1,\ldots,2n\}$\\
      4.\> \textbf{return} $\pi_J$
    \end{tabbing}
  \end{minipage}\par
  \noindent\makebox[.48\linewidth]{(a) Procedure \textsc{WeightedOrder}$(s,u)$}\hfill
  \makebox[.48\linewidth]{(b) Procedure \textsc{SampleOrder}$(s)$}\par
\end{pseudocode}

\paragraph{Weighted load in one round.}
\textsc{WeightedOrder} includes each point~$p$ in $A$ independently with probability~$s\mu_p$, capped at one, and lists $A$ first in an arbitrary fixed order.  It appends the remaining points one at a time, choosing $p$ from the remaining set~$M$ with probability~$u(p)/u(M)$.
With exact rejection sampling, \textsc{WeightedOrder} can run forever on bit streams for which a draw keeps rejecting.  With independent unbiased bits, it halts with probability one and runs in expected polynomial time.  Appendix~\ref{sec:random-sampling} gives the exact implementation.

To analyze the sampler, give each point~$p\in P$ an independent exponential ringing time~$\tau_p$ of rate~$\mu_p$, so that $\Pr(\tau_p>t)=e^{-\mu_p t}$ for $t\ge 0$.  Let $S_t=\{p\in P:\tau_p\le t\}$.  Let $\sigma$ be the clock order that lists $S_s$ first in a fixed order, followed by the remaining points in increasing order of their ringing times, breaking ties by the same fixed order.  Figure~\ref{fig:overview-clocks} illustrates how the edge test on $S_t\cup\{p,q\}$ relates the load to the edge count of $G(S_t)$.

To bound the edge count, consider the density of two-point traces on a nonempty set~$S$.  It is the number of distinct pairs that ranges isolate on $S$, divided by $|S|$.  For the restriction shown in Figure~\ref{fig:definition-traces}, the density is $2/3$.  The algorithm requires a bound~$\kappa$ on this density over all restrictions, as in the hypothesis of Ackerman, Keszegh, and P\'alv\"olgyi~\cite[Theorem~6]{ackerman-tangencies}.  Keller and Smorodinsky~\cite{keller} call this condition hereditary linearity.

\begin{definition*}[Hereditary sparsity]
A range space on $P$ satisfies the \emph{hereditary sparsity condition with parameter~$\kappa$} if
\begin{equation}\label{eq:sparse}
  |E(G(S))|\le \kappa |S| \qquad\text{for every } S\subseteq P.
\end{equation}
\end{definition*}

\begin{lemma}\label{lem:sampler}
Suppose the range space satisfies the hereditary sparsity condition with parameter~$\kappa$.  Fix $s\in\{1,\ldots,n\}$ and positive integer weights~$u(p)$.  For the pair~$(\pi,A)$ returned by \textsc{WeightedOrder}, we have $\E[|A|]\le s$ and
\begin{equation}\label{eq:sampler}
  \E[\mu\cdot L_{\pi,A}]\le 1+2\kappa(1+\ln(n/s)).
\end{equation}
\end{lemma}
\begin{proof}
We first bound the load for the clock order using sparsity to control the expected edge count, then couple the two orders so that moving to \textsc{WeightedOrder} can only decrease the load.

\textit{Clock bound.}  Since $\Pr[p\in S_t]=1-e^{-\mu_p t}\le\mu_p t$, summing over $p$ gives $\E[|S_t|]\le\min\{t,n\}$.

For distinct $p$ and $q$, let
\[
  h_{pq}(t)=\Pr\bigl[pq\in G(S_t\cup\{p,q\})\bigr].
\]
At fixed $t$, the event defining $h_{pq}(t)$ depends only on the other clocks, because the test set contains $p$ and $q$ regardless of when they ring.  The edge~$pq$ belongs to $G(S_t)$ exactly when $\tau_p\le t$, $\tau_q\le t$, and $pq\in G(S_t\cup\{p,q\})$, since $S_t\cup\{p,q\}=S_t$ once both endpoints have rung by $t$.  These three events depend on disjoint groups of independent clocks and are therefore mutually independent.  Summing the products of their probabilities over ordered pairs counts each edge twice and gives
\[
  \sum_{p\ne q}(1-e^{-\mu_p t})(1-e^{-\mu_q t})h_{pq}(t) =2\E[|E(G(S_t))|].
\]

With probability one, all ringing times are distinct.  On this event, if $s<\tau_p=t<\tau_q$, then $B_{\sigma}(p)=S_t$ and $G_{\sigma}(p,q)=G(S_t\cup\{q\})$.  Since $p\in S_t$, this is the graph in the definition of $h_{pq}(t)$, and $p$ contributes to $L_{\sigma,S_s}(q)$ exactly when the event defining $h_{pq}(t)$ occurs.  Integrating the product of this event probability, the density~$\mu_p e^{-\mu_p t}$, and the probability~$e^{-\mu_q t}$ that $\tau_q>t$ gives the expected contribution.  Weighting by $\mu_q$ and summing over ordered pairs gives the expected weighted contribution of the earlier points counted in the loads.  The unit terms contribute at most $\sum_{p\in P\setminus S_s}\mu_p \le \sum_{p\in P}\mu_p = 1$, so
\[
  \E[\mu\cdot L_{\sigma,S_s}]
  \le 1+\int_s^\infty a_\mu(t)\,dt, \qquad
  a_\mu(t)=\sum_{p\ne q}\mu_p\mu_q e^{-(\mu_p+\mu_q)t}h_{pq}(t).
\]

An exponential density decreases with time, so its value at $t>0$ is at most its average over $[0,t]$.  That average is the probability of ringing by $t$ divided by $t$, so
\[
  \mu_p e^{-\mu_p t}\le \frac{1-e^{-\mu_p t}}{t}.
\]
Use this bound for both endpoint densities in each summand of $a_\mu(t)$.  The edge-count identity then gives
\[
  a_\mu(t)\le \frac{1}{t^2}\sum_{p\ne q}(1-e^{-\mu_p t})(1-e^{-\mu_q t})h_{pq}(t)
  =\frac{2\E[|E(G(S_t))|]}{t^2}.
\]
For each realization of $S_t$, the hereditary sparsity condition \eqref{eq:sparse} gives $|E(G(S_t))|\le \kappa|S_t|$.  Taking expectations in this inequality and using $\E[|S_t|]\le\min\{t,n\}$ gives
\[
  a_\mu(t)\le \frac{2\kappa\E[|S_t|]}{t^2}
             \le \frac{2\kappa\min\{t,n\}}{t^2}.
\]
Integrating on $[s,n]$ and $[n,\infty)$ gives
\begin{equation}\label{eq:clock}
  \E[\mu\cdot L_{\sigma,S_s}]
  \le 1+2\kappa(1+\ln(n/s)).
\end{equation}

\textit{Coupling.}  Let $\alpha_p=\min\{1,s\mu_p\}$.  Since $\alpha_p\le s\mu_p$, we have $\E[|A|]\le s$.  Couple $A$ with $S_s$ using independent uniform variables~$U_p$ on $[0,1]$.  Include $p$ in $S_s$ if $U_p\le 1-e^{-s\mu_p}$ and in $A$ if $U_p\le\alpha_p$.  The inequality $\alpha_p\ge 1-e^{-s\mu_p}$ gives $S_s\subseteq A$.  Conditional on the $U_p$, sample the residual times~$\tau_p-s$ for $p\notin S_s$ as independent exponentials of rates~$\mu_p$, as prescribed by memorylessness.

Move $A\setminus S_s$ into the prefix and list~$A$ in the sampler's fixed order, preserving the relative order outside $A$.  Points in $A$ have zero load and do not contribute to the loads of other points.  For $p,q$ outside $A$ with $p$ before $q$, the prefix through $p$ can only grow.  Any range whose trace on the larger test set is $\{p, q\}$ has the same trace on the smaller test set, which contains both endpoints.  Thus no load increases, and \eqref{eq:clock} continues to apply.

Conditional on $A$, the residual times~$\tau_p-s$ for $p\notin A$ are independent exponentials of rates~$\mu_p$.  The next point from the remaining set~$M$ is $p$ with probability~$u(p)/u(M)$, exactly as in \textsc{WeightedOrder}.  This proves \eqref{eq:sampler}.
\end{proof}

\paragraph{Multiplicative weights across rounds.}
A point with small weight can contribute little to the weighted average even if its load makes a large contribution to the charging bound.  To control the total load on a hitting set, \textsc{SampleOrder} uses multiplicative weights~\cite{arora}.  It initializes all weights to $1$.  In each round, it calls \textsc{WeightedOrder} with the current weights, estimates the loads for the sampled pair~$(\pi,A)$, and multiplies each point's weight by a larger factor when that point has a larger load estimate.

We estimate the normalized loads from one point~$Q$ drawn uniformly from $P$, independently of the sampled pair~$(\pi,A)$ and of previous rounds.  If $Q\notin A$, test whether $pQ\in G_\pi(Q,p)$ for each $p$ after $Q$ in $\pi$.  This uses at most $n-1$ edge tests.  Set
\begin{equation}\label{eq:estimate}
  \widehat\ell(p)=
  \begin{cases}
    0, & \text{if }p\in A,\\
    1, & \text{if }p\notin A,\ Q\in B_\pi(p)\setminus A,\ \text{and }pQ\in G_\pi(Q,p),\\
    1/(n+1), & \text{otherwise}.
  \end{cases}
\end{equation}
For $p\notin A$, the estimate includes $1/(n+1)$ for $p$ itself and an additional $n/(n+1)$ if the point~$Q$ contributes to the load of $p$ in \eqref{eq:load}.  Each such point is drawn with probability~$1/n$, so $\E[\widehat\ell(p)\mid \pi,A]=L_{\pi,A}(p)/(n+1)$.

In round~$j$, let $(\pi_j,A_j)$ be the pair sampled with weights~$u_j$, and let $\widehat\ell_j$ be the resulting estimate.  The weight update is
\begin{equation}\label{eq:update}
  u_{j+1}(p)=(n+1)\left(1+\widehat\ell_j(p)\right)u_j(p).
\end{equation}
The factor~$n+1$ keeps the weights integral and cancels when they are normalized.  The three integer multipliers are $n+1$ for $p\in A_j$, $2(n+1)$ when the sampled point contributes to its load, and $n+2$ otherwise.

After $T=2n$ rounds, choose $J$ uniformly from $\{1,\ldots,T\}$ conditional on the completed rounds and return $\pi_J$ for use by the blind rule.  For each fixed set of at most $s$ points, this makes the expected total load of the returned order, measured with its prefix~$A_J$, equal to the expected average over the rounds.
The next lemma uses the per-round bound of Lemma~\ref{lem:sampler} to control the total load on a fixed hitting set.

\begin{lemma}[Cost for a fixed guess]\label{lem:fixed-guess}
Fix $s\in\{1,\ldots,n\}$.  On every fixed request sequence with offline optimum~$k \le s$, the blind rule with the order returned by \textsc{SampleOrder} has expected cost at most
  $(6+4\kappa)s(1+\ln(n/s))$.
\end{lemma}
\begin{proof}
Extend an optimal hitting set of size~$k$ for the fixed request sequence to a set~$S$ of size~$s$.  Write $\mu_j(p)=u_j(p)/u_j(P)$.  Normalizing the weight update in \eqref{eq:update} gives
\[
  \frac{\mu_{j+1}(p)}{\mu_j(p)}
  =\frac{1+\widehat\ell_j(p)}{1+\mu_j\cdot\widehat\ell_j}.
\]
Thus $p$ gains normalized weight exactly when its estimate~$\widehat\ell_j(p)$ exceeds the current weighted average~$\mu_j\cdot\widehat\ell_j$.  Both quantities lie in $[0,1]$.
Taking logarithms, applying $x/2 \le \ln(1+x) \le x$ on $[0,1]$, and summing 
over $j \in \{1, \dots, T\}$ with $\mu_1(p) = 1/n$ yields
$$
\ln(n\mu_{T+1}(p)) 
= \sum_{j=1}^T \ln(1 + \widehat{\ell}_j(p)) - \sum_{j=1}^T \ln(1 + \mu_j \cdot \widehat{\ell}_j)
\ge \frac{1}{2}\sum_{j=1}^T \widehat{\ell}_j(p) - \sum_{j=1}^T \mu_j \cdot \widehat{\ell}_j.
$$
Rearranging, averaging over $p \in S$, and applying the arithmetic--geometric 
mean inequality with $\sum_{p \in S} \mu_{T+1}(p) \le 1$ gives
\begin{equation}
\frac{1}{2s} \sum_{p \in S} \sum_{j=1}^T \widehat{\ell}_j(p) 
\le \frac{1}{s} \sum_{p \in S} \ln(n\mu_{T+1}(p)) + \sum_{j=1}^T \mu_j \cdot \widehat{\ell}_j 
\le \ln\frac{n}{s} + \sum_{j=1}^T \mu_j \cdot \widehat{\ell}_j. \label{eq:mwu-potential}
\end{equation}

Conditional on the preceding rounds, $\mu_j$ is fixed.  By \eqref{eq:estimate},
\[
  \E\left[(n+1)\widehat\ell_j(p)\,\middle|\,\pi_j,A_j,\text{preceding rounds}\right]=L_{\pi_j,A_j}(p).
\]
Averaging over $(\pi_j,A_j)$ and applying Lemma~\ref{lem:sampler} gives
\[
  \E\left[\mu_j\cdot\widehat\ell_j\,\middle|\,\text{preceding rounds}\right]
  \le \frac{1+2\kappa(1+\ln(n/s))}{n+1}.
\]
Take expectations in \eqref{eq:mwu-potential} and multiply by $2(n+1)/T$.  The conditional mean identity converts the left side to the mean load on $S$, averaged over the rounds.  Since $J$ is uniform and independent of the rounds, this gives
\[
  \frac{1}{s}\sum_{p\in S}\E[L_{\pi_J,A_J}(p)]
  \le \frac{2(n+1)}{T}\ln \left(\frac ns \right)+2+4\kappa \left(1+\ln \left(\frac ns \right)\right).
\]
Lemma~\ref{lem:charge} bounds the expected cost by $\E[|A_J|]$ plus the expected total load on $S$.  Using $\E[|A_J|]\le s$ from Lemma~\ref{lem:sampler} and $n+1\le T$ gives the desired cost bound.
\end{proof}

\subsection{Phase budgets and the online algorithm}

Recall that $n \ge 2$.  Since $k$ is unknown, the online algorithm uses the guesses~$s=1,2,4,\ldots$ with~$s<n$, one per phase.  During preprocessing, the calls to \textsc{SampleOrder} for different guesses use independent randomness.  For each guess~$s$, the budget~$B_s$ limits the number of distinct points the blind rule may select in that phase.  We take $B_s$ to be at least four times the preceding expected cost bound.  Let $s_{\max}$ be the largest power of two below $n$.  Since $s_{\max}\ge n/2$ and $B_{s_{\max}}\ge 24s_{\max}>n$, that phase has room for all $n$ points.

Fix a request sequence and a guess~$s$.  Condition on the sampled orders for all guesses smaller than~$s$.  If phase~$s$ is reached, these orders determine its starting request.  If $k\le s$, the suffix starting at that request has an offline optimum of at most $k\le s$.  By independence and Lemma~\ref{lem:fixed-guess}, the expected number of distinct first points on this suffix under $\pi_s$ is at most $B_s/4$.  A request starts the next phase when its first point under $\pi_s$ would be the $(B_s+1)$st distinct first point in phase~$s$.  Markov's inequality therefore bounds the probability of advancing by $1/4$ for each choice of earlier orders that reaches phase~$s$.  Averaging over those orders gives the same bound conditional on reaching phase~$s$.

Preprocessing takes the candidate set~$P$, the sparsity parameter~$\kappa$, and an oracle that tests adjacency in $G(S)$ for any $S\subseteq P$.  These inputs are shared by all calls to \textsc{SampleOrder}.

\begin{pseudocode}
  \begin{minipage}[t]{.48\linewidth}
    \hrule width\linewidth height0pt\relax
    \begin{tabbing}
      \pseudocodetitle{alg:online}{Online hitting set}
      \\[5mm]
      0.\quad\=1.1.\quad\=\quad\=\kill
      1.\> \textbf{for} $s=1,2,4,\ldots$ with $s<n$ \textbf{do}\\
      \> 1.1.\> $\pi_s\gets \text{\textsc{SampleOrder}}(s)$\\
      \> 1.2.\> $B_s\gets\lceil 4(6+4\kappa)\rceil\,s(1+\lceil\log_2(n/s)\rceil)$
    \end{tabbing}
  \end{minipage}\hfill
  \begin{minipage}[t]{.48\linewidth}
    \hrule width\linewidth height0pt\relax
    \begin{tabbing}
      0.\quad\=1.1.\quad\=\quad\=\kill
      1.\> $s\gets 1$, $X\gets\varnothing$, $X_1\gets\varnothing$\\
      2.\> \textbf{for} each range $R$ as it arrives \textbf{do}\\
      \> 2.1.\> $p\gets$ first point of $R$ in $\pi_s$\\
      \> 2.2.\> \textbf{if} $p\notin X_s$ and $|X_s|=B_s$ \textbf{then}\\
      \>\>\> $s\gets 2s$, $X_s\gets\varnothing$\\
      \>\>\> $p\gets$ first point of $R$ in $\pi_s$\\
      \> 2.3.\> $X_s\gets X_s\cup\{p\}$, $X\gets X\cup\{p\}$\\
      3. \> \textbf{return} $X$.
    \end{tabbing}
  \end{minipage}\par
  \noindent\makebox[.48\linewidth]{(a) Offline preprocessing}\hfill
  \makebox[.48\linewidth]{(b) Online selection}\par
\end{pseudocode}

The blind rule uses each request's first point even if $X$ already hits the request, and counts that point in $X_s$ even if an earlier phase selected it.  Thus, when $|X_s|=B_s$, step~2.2 can start a new phase with $p\in X\setminus X_s$.  At a phase change, keep $X$, initialize the next phase's set to empty, and process the request under its new order.

The hereditary sparsity condition \eqref{eq:sparse} gives the following guarantee.

\begin{theorem}\label{thm:sparse}
Suppose the range space satisfies the hereditary sparsity condition with parameter~$\kappa$.  On every fixed request sequence with offline optimum~$k\ge 1$, Algorithm~\ref{alg:online} has expected cost~$O((1+\kappa)k(1+\log(n/k)))$.  It receives each request as an explicit list of points of $P$.  Preprocessing uses $O(n^2\log n)$ queries to an oracle that decides whether a given pair is an edge of $G(S)$.  The rest of preprocessing takes expected polynomial time.
\end{theorem}
\begin{proof}
After each request, $X$ contains its selected point and all earlier selections, so it hits all requests seen so far.  Since $|X_{s_{\max}}|\le n<B_{s_{\max}}$, the phase-change condition cannot hold at $s_{\max}$.  If $k>s_{\max}$, the cost is at most $n\le 2s_{\max}<2k$.

For the remaining cost analysis, assume $k\le s_{\max}$ and let $s_*$ be the smallest guess at least $k$.  The number of points selected in phases with guesses below $s_*$ is at most the sum of their budgets.  If $s_*>1$, let $r$ be the largest guess below $s_*$.  For each guess~$s$, write $g_s=1+\lceil\log_2(n/s)\rceil$.  The earlier guesses have the form $r/2^i$ for integers~$i\ge 0$, with $g_{r/2^i}=g_r+i$, so their total budget is
\[
  O\!\left((1+\kappa)r\sum_{i\ge 0}2^{-i}(g_r+i)\right)
    =O((1+\kappa)r g_r).
\]
Since $r<k\le s_*\le 2r$, this is $O((1+\kappa)k(1+\log(n/k)))$.

The final $X$ is the union of the sets~$X_s$ at the ends of the reached phases, each of size at most~$B_s$.  By the conditional bound above, from $s_*$ onward the probability of reaching the next phase is at most one quarter of that of reaching the current phase.  Each successive budget is at most twice the preceding one, because the guess doubles and $g_s$ is nonincreasing.  The expected cost from $s_*$ onward is thus at most
\[
  \sum_{i\ge 0} B_{s_*}2^i4^{-i}=2B_{s_*}.
\]
Since $k\le s_*\le 2k$, this too is $O((1+\kappa)k(1+\log(n/k)))$.

Each round of \textsc{SampleOrder} makes at most $n-1$ edge tests.  The $T=2n$ rounds per guess and $O(\log n)$ guesses therefore use $O(n^2\log n)$ edge tests.  A weight is a product of $O(n)$ integers at most $2(n+1)$, so it has $O(n\log n)$ bits.  The required draws can be implemented exactly in expected polynomial time.
\end{proof}

Appendix~\ref{app:finite-sampling} gives a variant with the same expected-cost bound and polynomial worst-case preprocessing time apart from edge tests.

\paragraph{Arrival order and phase changes.}
For a fixed collection of requests assigned to a phase, the blind rule selects their distinct first points under the phase's order.
Use the ranges~$R_1=\{a,b\}$ and $R_2=\{b,c,d\}$ of Figure~\ref{fig:definition-traces} in an instance with at least $B_1-1$ candidates outside $\{a,b,c,d\}$.  Start with $B_1-1$ distinct singleton requests on these extra candidates, leaving room for one more distinct first point in the first phase.  Let the first phase order the four points as $\langle a,b,c,d\rangle$ and put $b$ first among them in the second phase.  If $R_1$ precedes $R_2$, the algorithm selects $a$ for $R_1$, then switches phases on $R_2$ and selects $b$.  In the reverse order, the algorithm selects $b$ for $R_2$, then switches phases on $R_1$ and selects $b$ again.  Thus the two arrival orders select $\{a,b\}$ and $\{b\}$, respectively, among the four points.
Indeed, a later occurrence of the same range can add a point if its first point in the new order has yet to be selected.

\begin{remark}
A variant adds the first point~$p$ to $X$ only if the arriving range satisfies $X\cap R=\varnothing$.
In Figure~\ref{fig:definition-traces}, take $\pi=\langle a,b,c,d\rangle$.  The blind rule selects $a$ for $R_1$ and $b$ for $R_2$, so its final set is $\{a,b\}$ in either arrival order.  If $R_1$ arrives first, the rule that skips already hit ranges selects $a$ and then $b$, since $a$ does not hit $R_2$.  If $R_2$ arrives first, that rule selects only $b$, since $b$ also hits $R_1$.

The variant still records every first point in $X_s$ and switches phases based on $X_s$.  The points added to $X$ are then a subset of those selected by Algorithm~\ref{alg:online}, so the same cost bound applies.  For explicit ranges, the test~$X\cap R=\varnothing$ can be done while scanning $R$ for its first point.

For disks, the test reduces to a nearest-neighbor query on $X$.  Applying the deamortized logarithmic method of Overmars and van Leeuwen~\cite{overmars-vanleeuwen} to static Voronoi structures gives $O(\log^2 n)$ worst-case time for the test and each insertion.  We omit the implementation details.
\end{remark}

\section{Online hitting set for planar disks}\label{sec:disks}

We prove Theorem~\ref{thm:disk} by verifying the hereditary sparsity condition~\eqref{eq:sparse}, then show how to test Delaunay adjacency and find the first point of an arriving disk in the current order.  We count arithmetic operations, square-root computations, and comparisons on coordinates at unit cost.  Appendix~\ref{sec:random-sampling} accounts separately for random-bit sampling and integer-weight arithmetic.

\begin{proof}[Proof of Theorem~\ref{thm:disk}]
Delaunay graphs of disks admit plane straight-line drawings~\cite[Section~2.1.1]{ackerman}.  To see this, fix $S\subseteq P$.  For each edge of $G(S)$, choose a closed disk whose trace on $S$ consists of the two endpoints and draw the edge as their straight-line segment.

For two edges with disjoint endpoints, write their disks as $f(x)\le 0$ and $g(x)\le 0$, where $f(x)=\|x-c\|^2-\rho^2$ for a disk with center~$c$ and radius~$\rho$, and $g$ is defined analogously.  The quadratic terms cancel in $f-g$, so it is affine.  It is strictly negative at the endpoints of the first segment and strictly positive at the endpoints of the second.  It therefore has opposite strict signs along the two segments, which are disjoint.

By convexity each segment lies in its disk, so it contains no third point of $S$.  Edges sharing an endpoint can therefore meet only at that endpoint.  The resulting plane graph has $|E(G(S))|\le 3|S|$, giving \eqref{eq:sparse} with $\kappa=3$.  Theorem~\ref{thm:sparse} gives the expected cost bound.
\end{proof}

\subsection{Delaunay edge tests via geometric inversion}\label{sec:disk-implementation}

During preprocessing, \textsc{SampleOrder} uses the edge tests in \eqref{eq:estimate} to compute the normalized load estimates for its multiplicative weight updates.  These queries concern all possible disks and are made before any requests arrive.

In a round with $Q=p\notin A$, let $T_p=B_\pi(p)\setminus\{p\}$ be the points strictly before $p$.  For each $q$ after $p$ in $\pi$, the edge test asks whether a disk contains $p,q$ and misses $T_p$.  Geometric inversion about $p$ turns this into a convex-hull membership test.

\begin{lemma}[Disk witnesses via inversion]\label{lem:disk-inversion}
Let $p,q$ be distinct points in the plane, and let $T_p$ be a finite set disjoint from $\{p,q\}$.  Define
\[
  a_p(r)=\frac{2(r-p)}{\|r-p\|^2}\quad (r\ne p),
  \qquad
  H_p=\conv\bigl(\{0\}\cup\{a_p(r):r\in T_p\}\bigr).
\]
There is a closed disk containing $p,q$ and missing $T_p$ if and only if $a_p(q)\notin H_p$.  When such a disk exists, it can be chosen with both $p$ and $q$ on its boundary.
\end{lemma}

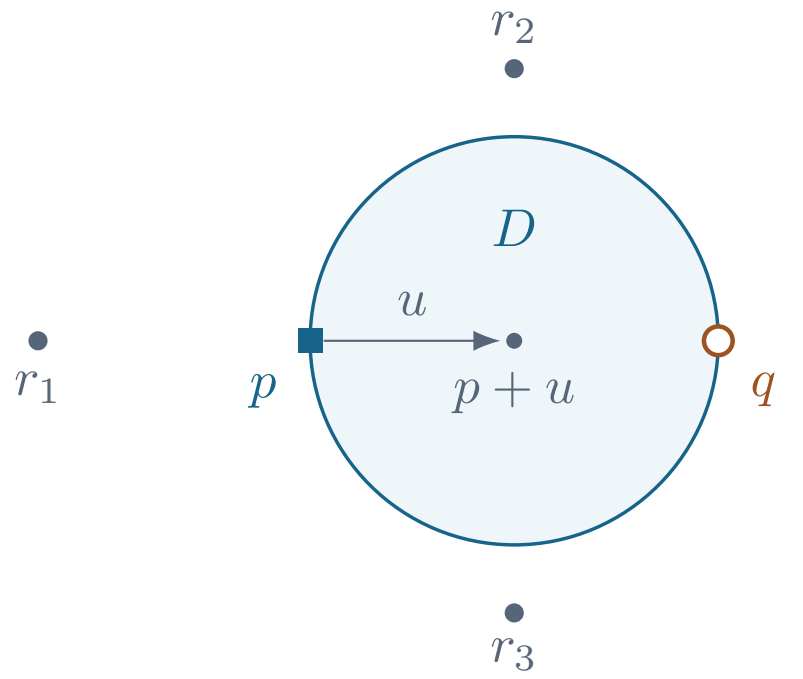
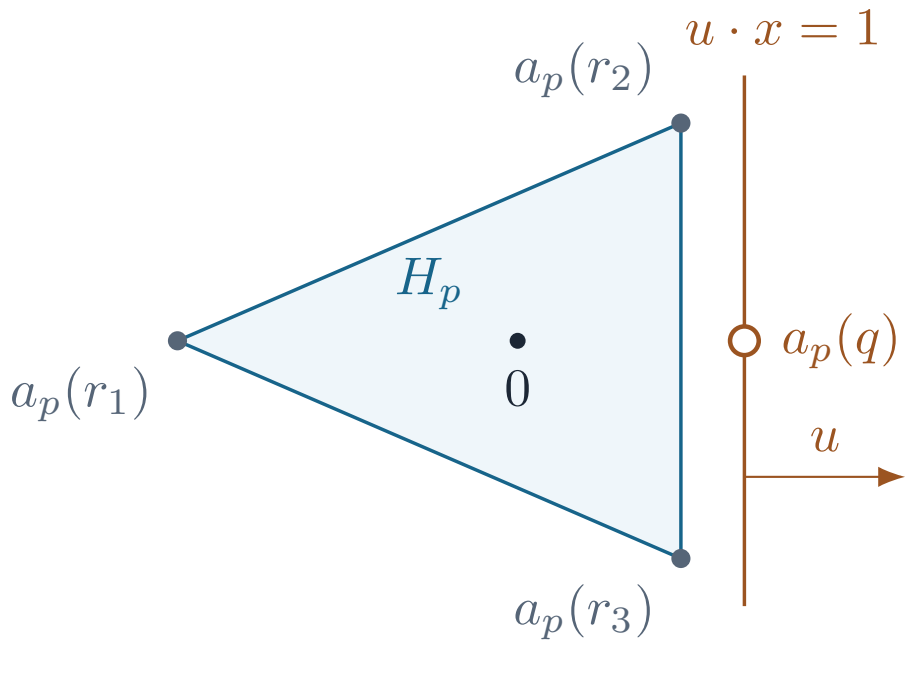
\begin{figure}[!htbp]
  \centering
  \begin{subfigure}[t]{.47\linewidth}
    \centering
    \begin{tikzpicture}[x=.9cm,y=.9cm,font=\small,text=overviewink]
      \useasboundingbox (-2.5,-2.5) rectangle (3.55,2.55);
      \draw[overviewblue,line width=.65pt,fill=overviewfill] (1.5,0) circle[radius=1.5];
      \node[text=overviewblue] at (1.5,.83) {$D$};
      \draw[overviewslate,-{Latex[length=1.8mm]}] (.1,0)
        -- node[above=1pt] {$u$} (1.4,0);
      \fill[overviewslate] (1.5,0) circle[radius=1.5pt] node[below=2pt] {$p+u$};
      \fill[overviewslate] (-2,0) circle[radius=1.8pt] node[below=2pt] {$r_1$};
      \fill[overviewslate] (1.5,2) circle[radius=1.8pt] node[above=1pt] {$r_2$};
      \fill[overviewslate] (1.5,-2) circle[radius=1.8pt] node[below=1pt] {$r_3$};
      \node[fill=overviewblue,minimum size=4.6pt,inner sep=0pt,
        label={[text=overviewblue]below left:$p$}] at (0,0) {};
      \node[circle,draw=overvieworange,line width=.85pt,fill=white,minimum size=5.4pt,
        inner sep=0pt,label={[text=overvieworange]below right:$q$}] at (3,0) {};
    \end{tikzpicture}
    \caption{A disk through $p$ and $q$ that misses $T_p$.}
    \label{fig:disk-inversion-a}
  \end{subfigure}\hfill
  \begin{subfigure}[t]{.47\linewidth}
    \centering
    \begin{tikzpicture}[x=.9cm,y=.9cm,font=\small,text=overviewink]
      \useasboundingbox (-3.5,-2.5) rectangle (3.6,2.55);
\draw[overviewblue,line width=.65pt,fill=overviewfill]
        (-2.5,0) -- (1.2,1.6) -- (1.2,-1.6) -- cycle;
      \node[text=overviewblue] at (-.65,.42) {$H_p$};
      \fill[overviewslate] (-2.5,0) circle[radius=1.8pt] node[below left=1pt] {$a_p(r_1)$};
      \fill[overviewslate] (1.2,1.6) circle[radius=1.8pt] node[above left=1pt] {$a_p(r_2)$};
      \fill[overviewslate] (1.2,-1.6) circle[radius=1.8pt] node[below left=1pt] {$a_p(r_3)$};
      \fill[overviewink] (0,0) circle[radius=1.5pt] node[below=2pt] {$0$};
      \draw[overvieworange,line width=.65pt] ({5/3},-1.95)
        -- ({5/3},1.95);
      \node[text=overvieworange,above=2pt] at (1.95,1.95) {$u\cdot x=1$};
      \draw[overvieworange,-{Latex[length=1.8mm]}] ({5/3},-1)
        -- node[above=1pt] {$u$} (2.85,-1);
      \node[circle,draw=overvieworange,line width=.85pt,fill=white,minimum size=5.4pt,
        inner sep=0pt,label={[text=overvieworange]right:$a_p(q)$}] at ({5/3},0) {};
    \end{tikzpicture}
    \caption{The line through $a_p(q)$ separates it from $H_p$.}
    \label{fig:disk-inversion-b}
  \end{subfigure}
  \caption{Inversion at $p$. The disk with center~$p+u$ and radius~$\|u\|$ maps, away from $p$, to the halfplane~$u\cdot x\ge 1$. The hull~$H_p$ lies strictly outside this halfplane.}
  \label{fig:disk-inversion}
\end{figure}

With $p$ and $T_p$ fixed in a sampling round, Lemma~\ref{lem:disk-inversion} gives
\[
  pq\in G_\pi(p,q) \quad\Longleftrightarrow\quad a_p(q)\notin H_p.
\]
Thus all edge tests in the round use the same hull~$H_p$.  In Figure~\ref{fig:disk-inversion}, inversion maps the witness disk with $p$ removed to a closed halfplane that misses the hull.

\begin{proof}[Proof of Lemma~\ref{lem:disk-inversion}]
Suppose a disk with center~$p+u$ and radius~$\rho$ contains $p,q$ and misses $T_p$.  Since the disk contains $p$, we have $b=\rho^2-\|u\|^2\ge 0$.  For $r\ne p$, expanding the disk inequality gives $2u\cdot(r-p)+b\ge\|r-p\|^2$.  Since $\|a_p(r)\|^2=4/\|r-p\|^2$, division by $\|r-p\|^2$ shows that $r$ belongs to the disk exactly when
\[
  u\cdot a_p(r)+\tfrac b4\|a_p(r)\|^2\ge 1.
\]
Because $b\ge 0$, the left-hand side is convex in the transformed point.  It is zero at the origin and less than $1$ at $a_p(r)$ for every $r\in T_p$, so it is less than $1$ throughout $H_p$.  At $a_p(q)$ the value is at least $1$, so $a_p(q)$ lies outside $H_p$.

Conversely, if $a_p(q)\notin H_p$, take a normal vector~$v$ to a line that strictly separates $a_p(q)$ from $H_p$, oriented so that
\[
  a_p(q)\cdot v>\max_{x\in H_p}x\cdot v\ge 0.
\]
Set~$u=v/(a_p(q)\cdot v)$, whose denominator is positive since $0\in H_p$.  The disk with center~$p+u$ and radius~$\|u\|$ has $b=0$ and $u\cdot a_p(q)=1$, so both $p$ and $q$ lie on its boundary.  Every $r\in T_p$ satisfies $u\cdot a_p(r)<1$ and hence lies outside the disk.
\end{proof}

For each $p\in P$, precompute $a_p(r)$ for all $r\in P\setminus\{p\}$ and sort these points together with the origin by $x$-coordinate, breaking ties by $y$-coordinate.  The resulting list depends only on $P$ and $p$ and is reused across all rounds and guesses.

In a round with $Q=p\notin A$, construct the upper and lower hull chains of $H_p$ from the origin and the images of $T_p$, using their order in the stored list.  Then scan the query images~$a_p(q)$, for $q$ after $p$ in $\pi$, in the same order while advancing along the two chains.  At each query image's $x$-coordinate, the chains give the lowest and highest points of $H_p$.  The image belongs to $H_p$ exactly when its $y$-coordinate lies between their $y$-coordinates, including equality; if the hull has no point at that $x$-coordinate, the image lies outside.  If $H_p$ is a point or a segment, test membership directly.  Each chain is traversed at most once, so the round takes $O(n)$ geometric operations.

Sorting all $n$ lists takes $O(n^2\log n)$ geometric operations.  Together with the $O(n)$ rounds for each of the $O(\log n)$ guesses, the total is $O(n^2\log n)$ geometric operations.  On rational inputs, these computations use only rational arithmetic.  The edge tests provide the estimates in \eqref{eq:estimate} for the weight updates in \eqref{eq:update}, allowing \textsc{SampleOrder} to generate the priority orders before any requests arrive.

\subsection{First-point queries for disks}\label{sec:disk-first-point}
For each order, build a balanced binary tree whose leaves are the candidate points in that order.  A nearest-neighbor query in the left subtree, with the disk center as the query point, tells us whether any candidate in that subtree lies in the disk.  Descend left when the subtree meets the disk and right otherwise.  Voronoi point location answers each nearest-neighbor query in $O(\log n)$ time, and the tree search takes $O(\log^2 n)$ worst-case time.  Constructing the structures for all $O(\log n)$ orders takes $O(n\log^3 n)$ geometric operations, within the $O(n^2\log n)$ preprocessing bound.  Each request requires at most two searches and $O(1)$ additional operations.  Appendix~\ref{sec:disk-order-queries} gives the construction and its analysis.

\section{Shortest-path balls in minor-free graphs}\label{sec:graph}

Let $H$ be a finite connected undirected graph with positive edge lengths, and let $d_H$ denote its shortest-path distance.  For $c\in V(H)$ and $\rho\ge 0$, write
\[
  B_H(c,\rho)=\{x\in V(H):d_H(c,x)\le \rho\}.
\]
A ball meeting $P$ induces the request~$B_H(c,\rho)\cap P$.  We first obtain \eqref{eq:sparse} by showing that each $G(S)$ is a subgraph of a minor of $H$, then describe the edge tests and request handling.

\subsection{Voronoi cells and the sparsity bound}

For unweighted graphs, the minor containment follows by applying the result in~\cite[Lemma~2]{bousquet} to singleton balls.  A related contraction of Voronoi cells appears in the support construction of Browne and Chang~\cite[Section~3]{browne}.  We give the argument for positive edge lengths.

Fix a nonempty $S\subseteq P$.  Assign each vertex of $H$ to its nearest site in $S$, breaking ties by a fixed order on $S$.  Let $x$ be a vertex assigned to $p$ and let $y$ lie on a shortest path from $x$ to $p$.  For each $a\in S$, the triangle inequality and $d_H(x,p)=d_H(x,y)+d_H(y,p)$ give
\[
  d_H(y,a)-d_H(y,p)\ge d_H(x,a)-d_H(x,p)\ge 0.
\]
Thus $p$ remains nearest at $y$, and any site tied with $p$ at $y$ was also tied at $x$.  The fixed tie-breaking order therefore assigns $y$ to $p$, so each cell is connected.  Contract the cells, delete loops, and merge parallel edges to obtain a simple minor~$H_S$ with vertex set~$S$.

Let $B_H(c,\rho)$ have trace~$\{p,q\}$ on $S$.  By the triangle inequality, every vertex on a shortest path from $c$ to $p$ is closer to $p$ than to any site in $S\setminus\{p,q\}$; the analogous statement holds on a shortest path from $c$ to $q$.  Both paths therefore lie in the union of the cells of $p$ and $q$.  Together they connect $p$ to $q$, so these cells are adjacent.  Thus $pq\in H_S$, and $G(S)$ is a subgraph of $H_S$.

For a ball centered in an edge interior, the same sparsity argument applies after subdividing the edge at the center while preserving its length.  The original vertices keep their cell assignments, and the new vertex shares a cell with a neighbor on a shortest path to its assigned site.  Thus subdivision gives the same cell adjacency graph.

If $H$ is planar, then $H_S$ is planar and $|E(G(S))|\le 3|S|$.  If $H$ excludes a fixed graph~$F$ as a minor, then so does $H_S$, and Mader's theorem~\cite{mader} gives $|E(G(S))|=O_F(|S|)$.  Hence \eqref{eq:sparse} holds with $\kappa=O_F(1)$, and Theorem~\ref{thm:sparse} gives the expected cost bound in Theorem~\ref{thm:graph}.

\subsection{Shortest-path preprocessing and request handling}

Precompute the distances from each candidate to all vertices of $H$ using Dijkstra's algorithm.  To test whether $pq\in G(S)$ for distinct $p,q\in S$, consider each vertex~$c$ as a center and use the smallest radius for which the ball contains both $p$ and $q$.  This ball has trace~$\{p,q\}$ on $S$ exactly when
\[
  \max\{d_H(c,p),d_H(c,q)\}
  <
  \min_{a\in S\setminus\{p,q\}}d_H(c,a),
\]
where $\min\varnothing=+\infty$.  The inequality is strict because a vertex at distance equal to the radius belongs to the ball.

When a ball~$B_H(c,\rho)$ arrives, compare $\rho$ with the stored distances from $c$ to the candidates and pass the resulting range to Algorithm~\ref{alg:online}.  For rational edge lengths, each distance has polynomial bit length, since it is a sum of edge lengths along a simple path.  Constructing the distance table, testing an edge, and processing a request with a rational radius therefore take polynomial time in their input bit lengths.

Sparsity of the host graph alone is not sufficient.  Appendix~\ref{sec:degree3} shows that even equal-radius balls in graphs of maximum degree at most three can realize arbitrary range families.

\section{Matching lower bounds via interval hitting}\label{sec:lower}

We first prove the lower bound for interval requests, each of which is a nonempty consecutive block of an ordered candidate set.  We then realize the construction by disks and by shortest-path balls on a path.

In the offline setting, the right-endpoint greedy algorithm is optimal, and the minimum hitting-set size equals the maximum number of pairwise disjoint intervals (Gallai's theorem, see~\cite[Theorem~1.1(a)]{ahz}).  For online interval hitting, Even and Smorodinsky~\cite[Proposition~6]{even} proved the exact deterministic competitive ratio~$\lfloor\log_2 n\rfloor+1$.  The one-dimensional case of the construction of De, Mandal, and Singh~\cite{dms-polygons} gives a logarithmic lower bound for randomized interval hitting on integer candidates with offline optimum one.

\begin{theorem}[Interval hitting]\label{thm:interval-lower}
For all integers~$n,k$ with $1\le k\le n$, there is an ordered set~$P$ of $n$ candidates such that, for every randomized online algorithm on $P$, some fixed sequence of interval requests with offline optimum~$k$ forces the algorithm to select at least $k(1+\tfrac12\lfloor\log_2(n/k)\rfloor)$ points in expectation.
\end{theorem}

\begin{proof}
Put $h=\lfloor\log_2(n/k)\rfloor$.  Partition the first $k2^h$ candidates into $k$ consecutive blocks of $2^h$ candidates each, and let $I_0^j$ be the block for group~$j$, where $1\le j\le k$.  The remaining candidates belong to no request.

For each group~$j$ and each $i=1,\ldots,h$, split $I_{i-1}^j$ into two equal consecutive halves and choose $I_i^j$ by a fair coin flip.  Toss all the coins before the requests arrive, independently of one another and of the algorithm's random choices.  Process the groups in order, revealing $I_0^j,\ldots,I_h^j$ in order within group~$j$.  The sole point of $I_h^j$ hits every request in group~$j$, and the initial intervals~$I_0^1,\ldots,I_0^k$ are disjoint.  Thus every sequence in the distribution has offline optimum~$k$.  Figure~\ref{fig:dyadic-blocks-a} shows the initial intervals, and Figure~\ref{fig:dyadic-blocks-b} shows one realization of the nested requests for group~$1$.

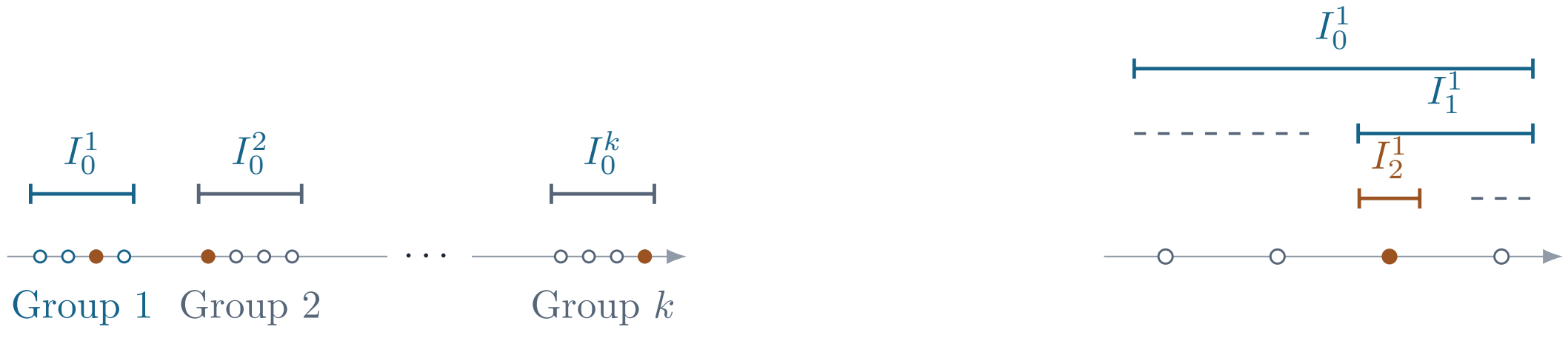
\begin{figure}[!htbp]
  \centering
  \begin{subfigure}[t]{.47\linewidth}
    \centering
    \begin{tikzpicture}[font=\small,text=overviewink]
      \useasboundingbox (-3.2,-.82) rectangle (3.35,2.15);
      \draw[overviewslate!65,-{Latex[length=1.8mm]}] (-3.02,0) -- (3.05,0);
      \foreach \sitecenter/\groupcolor/\i/\survivor in {-2.35/overviewblue/1/.125,-.85/overviewslate/2/-.375,2.3/overviewslate/k/.375} {
        \draw[\groupcolor,line width=.9pt]
          ({\sitecenter-.46},.56) -- ({\sitecenter+.46},.56) node[midway,above=1pt,text=overviewblue] {$I_0^\i$}
          ({\sitecenter-.46},.47) -- ({\sitecenter-.46},.65)
          ({\sitecenter+.46},.47) -- ({\sitecenter+.46},.65);
        \foreach \offset in {-.375,-.125,.125,.375} {
          \draw[\groupcolor,fill=white,line width=.55pt]
            ({\sitecenter+\offset},0) circle[radius=1.5pt];
        }
        \fill[overvieworange] ({\sitecenter+\survivor},0) circle[radius=1.8pt];
      }
      \node[fill=white,inner xsep=4pt,inner ysep=1pt] at (.75,0) {$\cdots$};
      \node[text=overviewblue] at (-2.35,-.46) {Group $1$};
      \node[text=overviewslate] at (-.85,-.46) {Group $2$};
      \node[text=overviewslate] at (2.3,-.46) {Group $k$};
    \end{tikzpicture}
    \caption{Disjoint initial intervals on the $k$ groups.}
    \label{fig:dyadic-blocks-a}
  \end{subfigure}\hfill
  \begin{subfigure}[t]{.47\linewidth}
    \centering
    \begin{tikzpicture}[font=\small,text=overviewink]
      \useasboundingbox (-2.3,-.82) rectangle (2.3,2.15);
      \draw[overviewslate!65,-{Latex[length=1.8mm]}] (-2.05,0) -- (2.05,0);
      \foreach \point in {-1.5,-.5,1.5} {
        \draw[overviewslate,fill=white,line width=.55pt]
          (\point,0) circle[radius=1.8pt];
      }
      \fill[overvieworange] (.5,0) circle[radius=2pt];
      \draw[overviewblue,line width=.9pt]
        (-1.78,1.68) -- (1.78,1.68) node[midway,above=1pt,text=overviewblue] {$I_0^1$}
        (-1.78,1.59) -- (-1.78,1.77) (1.78,1.59) -- (1.78,1.77);
      \draw[overviewslate,dashed,line width=.7pt] (-1.78,1.10) -- (-.22,1.10);
      \draw[overviewblue,line width=.9pt]
        (.22,1.10) -- (1.78,1.10) node[midway,above=1pt,text=overviewblue] {$I_1^1$}
        (.22,1.01) -- (.22,1.19) (1.78,1.01) -- (1.78,1.19);
      \draw[overviewslate,dashed,line width=.7pt] (1.23,.52) -- (1.77,.52);
      \draw[overvieworange,line width=.9pt]
        (.23,.52) -- (.77,.52) node[midway,above=1pt,text=overvieworange] {$I_2^1$}
        (.23,.43) -- (.23,.61) (.77,.43) -- (.77,.61);
    \end{tikzpicture}
    \caption{One revealed chain for group~$1$ with $h=2$.}
    \label{fig:dyadic-blocks-b}
  \end{subfigure}
  \caption{The Matryoshka-doll construction.  In group~$1$, the solid bars show the revealed intervals~$I_0^1\supset I_1^1\supset I_2^1$, and the dashed bars show the other half at each split.  Each half is chosen with probability~$1/2$.  Each orange point represents the final surviving candidate in its group and hits all requests there.}
  \label{fig:dyadic-blocks}
\end{figure}

Fix a randomized algorithm.  Let $H_i^j$ count the selected points in $I_i^j$ at the end of that request.  For $i\ge 1$, let $A_i^j$ count the points in $I_i^j$ first selected while processing that request.  Set~$A_0^j=H_0^j$ to account for any points of group~$j$ selected before its first request as well.  For $i\ge 1$, condition on the revealed requests and selections through $I_{i-1}^j$, including those in earlier groups.  The next coin is independent of this history, so each of the $H_{i-1}^j$ selected points in $I_{i-1}^j$ belongs to $I_i^j$ with probability~$1/2$.  Each point counted by $H_i^j$ either was already selected and survived the split or was first selected while processing $I_i^j$.  By linearity of expectation,
\[
  \E[H_i^j]=\tfrac12\E[H_{i-1}^j]+\E[A_i^j],
\]
where the expectation is over the interval choices and the algorithm.  Every request must be hit, so $H_i^j\ge 1$.  Summing the recurrence within group~$j$ gives
\[
  \sum_{i=0}^{h}\E[A_i^j]
  =\E[H_h^j]+\frac12\sum_{i=0}^{h-1}\E[H_i^j]
  \ge 1+\frac h2.
\]
Each selected point is counted by at most one $A_i^j$, since each group counts distinct selections and the groups are disjoint.  Summing over $j=1,\ldots,k$ gives expected cost at least $k(1+h/2)$ over both the sequence and the algorithm.  Hence some fixed sequence has expected cost at least $k(1+h/2)$ over the algorithm's random choices.
\end{proof}

In Appendix~\ref{app:refined-intervals}, we show that the leading factor $1/2$ in Theorem~\ref{thm:interval-lower} is tight for online interval hitting: a randomly shifted rank rule achieves the exact worst-case expected cost $1 + \frac{h}{2} + \frac{n - 2^h}{2^{h+1}}$ when $k = 1$ (with $h = \lfloor\log_2 n\rfloor$), and a recursive splitting algorithm achieves expected cost $\frac{1}{2}k\log_2(n/k) + O\bigl(k\sqrt{1 + \log(n/k)}\bigr)$ when $k$ is known, as well as a competitive ratio of $\frac{1}{2}\log_2 n + O(\sqrt{\log n})$ without knowledge of $k$.

\begin{proof}[Proof of Theorem~\ref{thm:disk-lower}]
Put $h=\lfloor\log_2(n/k)\rfloor$.  Place the $k$ groups of $2^h$ points at unit spacing on the horizontal axis.  For a requested interval~$I_i^j$, let $a,b$ be the coordinates of its leftmost and rightmost points.  Use the disk centered at $((a+b)/2,0)$ with radius~$(b-a)/2+1/4$.  Along the axis it extends $1/4$ past each endpoint, so it contains exactly the points of the block within its group.  Moving an endpoint inward by $\delta$ moves the center by $\delta/2$ and reduces the radius by the same amount.  Thus the disk for a subblock lies inside the disk for the full block.  Separate the groups so that the disks containing the full groups are disjoint, and place any remaining candidates outside these disks.  The disk requests have the same traces on the candidates as the interval requests in Theorem~\ref{thm:interval-lower}, giving the stated lower bound.
\end{proof}

\begin{corollary}[Shortest-path balls on a path]\label{cor:path-lower}
For all integers~$n,k$ with $1\le k\le n$, there is a path~$H$ with unit edge lengths and a set~$P\subseteq V(H)$ of $n$ candidate vertices such that, for every randomized online algorithm with candidate set~$P$, some fixed sequence of shortest-path ball requests with offline optimum~$k$ forces the algorithm to select at least $k(1+\tfrac12\lfloor\log_2(n/k)\rfloor)$ vertices in expectation.
\end{corollary}

\begin{proof}
Realize the groups from the proof of Theorem~\ref{thm:interval-lower} on a path with unit edge lengths, using even-indexed vertices as candidates.  Place the groups consecutively and any extra candidates after the last group.  For an interval~$I_i^j$ with endpoint indices~$a,b$, its midpoint~$(a+b)/2$ is a vertex, and the ball centered there with radius~$(b-a)/2$ contains precisely the candidates in that block.  The request traces are unchanged, giving the same lower bound.
\end{proof}

\paragraph{AI Statement}
The authors used ChatGPT 5.6 sol and ChatGPT 6.1 sol in technical discussions that contributed to the development of the approach combining exponential clocks and multiplicative weights.  These models also assisted with language editing and manuscript presentation.  The authors take full responsibility for the mathematical results, proofs, and final manuscript.

\paragraph{Acknowledgment.} Zhao Zhang thanks Yingli Ran and Chunying Ren for insightful discussions early in the project.

\appendix
\section{Sampling with random bits}\label{sec:random-sampling}

To draw a uniform integer in $\{0,\ldots,Z-1\}$, generate $\lceil\log_2 Z\rceil$ independent unbiased bits and repeat with fresh bits if the resulting integer is at least $Z$.  This rejection sampler~\cite{vonneumann1951} has success probability at least $1/2$ per attempt.  Its output is exact and it uses at most two attempts in expectation.

For each prefix decision in \textsc{WeightedOrder}, draw an independent uniform integer from $\{0,\ldots,u(P)-1\}$ and include $p$ if the draw is below $su(p)$.  For a weighted draw from $M$, partition $[0,u(M))$ into consecutive half-open intervals of lengths~$u(p)$ in the fixed order of points in $M$.  A uniform integer in $[0,u(M))$ lies in the interval for $p$ with probability~$u(p)/u(M)$, and a scan of the cumulative weights identifies the selected point.  The draws of $Q$ and $J$ are uniform integer draws as well.  There are polynomially many draws and each weight has $O(n\log n)$ bits, so sampling and weight arithmetic take expected polynomial time.

\subsection{Sampling in polynomial worst-case time}\label{app:finite-sampling}
For a polynomial worst-case bound on preprocessing, let $N$ be the smallest power of two at least $n$.  \textsc{FiniteWeightedOrder} rounds the prefix probabilities up to multiples of $1/N^2$ and allows at most $2\log_2 N$ rejection attempts at each position.  If all fail, it appends the first remaining point in the fixed order.

\begin{pseudocode}
  \textbf{Procedure} \textsc{FiniteWeightedOrder}$(s,N,u)$\par
  \begin{tabbing}
    0.\quad\=1.1.\quad\=\quad\=\kill
    1.\> $A\gets\varnothing$\\
    2.\> \textbf{for} each $p\in P$ \textbf{do}\\
    \> 2.1.\> $b_p\gets\left\lceil N^2\min\{1,su(p)/u(P)\}\right\rceil$\\
    \> 2.2.\> Draw $v$ independently and uniformly from $\{0,\ldots,N^2-1\}$\\
    \> 2.3.\> \textbf{if} $v<b_p$ \textbf{then} add $p$ to $A$\\
    3.\> Start $\pi$ with the points of $A$ in a fixed order and set $M\gets P\setminus A$\\
    4.\> \textbf{while} $M\ne\varnothing$ \textbf{do}\\
    \> 4.1.\> Draw at most $2\log_2 N$ independent uniform $\lceil\log_2 u(M)\rceil$-bit integers,\\
    \>\> stopping at the first $v<u(M)$ and accepting it\\
    \> 4.2.\> \textbf{if} an integer $v$ was accepted \textbf{then}\\
    \>\>\> Take $q\in M$ whose cumulative weight interval contains $v$\\
    \>\> \textbf{else}\\
    \>\>\> Take the first remaining point $q\in M$ in the fixed order\\
    \> 4.3.\> Append $q$ to $\pi$ and remove it from $M$\\
    5.\> \textbf{return} $(\pi,A)$
  \end{tabbing}
\end{pseudocode}

In \textsc{SampleOrder}, use this sampler and run $T=2N$ rounds.  In each round, draw $Q$ uniformly from the $n$ points and $N-n$ dummy positions, independently of $(\pi,A)$ and previous rounds.  Set
\begin{equation}\label{eq:finite-estimate}
  \widehat\ell^{\mathrm{fin}}(p)=
  \begin{cases}
    0, & \text{if }p\in A,\\
    1, & \text{if }p\notin A,\ Q\in B_\pi(p)\setminus A,\ \text{and }pQ\in G_\pi(Q,p),\\
    1/(N+1), & \text{otherwise}.
  \end{cases}
\end{equation}
If $Q$ is a dummy position, every estimate outside $A$ is $1/(N+1)$.  Use the update \eqref{eq:update} with $N+1$ in place of $n+1$ and $\widehat\ell_j^{\mathrm{fin}}$ in place of $\widehat\ell_j$.  After the rounds, draw $J$ uniformly from $\{1,\ldots,T\}$ and return $\pi_J$.

\begin{lemma}\label{lem:finite-sampler}
Suppose the range space satisfies the hereditary sparsity condition with parameter~$\kappa$.  Fix $s\in\{1,\ldots,n\}$ and positive integer weights~$u(p)$, and let $N$ be the smallest power of two at least $n$.  For the pair~$(\pi,A)$ returned by \textsc{FiniteWeightedOrder}, with $\mu_p=u(p)/u(P)$, we have
\[
  \E[|A|]\le s+1,\qquad
  \E[\mu\cdot L_{\pi,A}]\le 2+2\kappa(1+\ln(n/s)).
\]
The resulting variant of \textsc{SampleOrder} satisfies the cost bound in Lemma~\ref{lem:fixed-guess} and has polynomial worst-case running time apart from edge tests.
\end{lemma}
\begin{proof}
Let $\alpha_p$ be the rounded inclusion probability.  Since $\alpha_p\le s\mu_p+1/N^2$, we have $\E[|A|]\le s+n/N^2\le s+1$.  The prefix coupling in the proof of Lemma~\ref{lem:sampler} still applies because $\alpha_p\ge\min\{1,s\mu_p\}$.  With exact weighted sampling after this prefix, the load of each point is no greater than in the coupled clock order, so \eqref{eq:clock} applies.

Couple \textsc{FiniteWeightedOrder} with a sampler that uses unlimited rejection attempts, keeping the prefix and draws identical until all $2\log_2 N$ attempts at some position fail.  At each position, conditional on acceptance within the limit at all earlier positions, this failure has probability at most $1/N^2$.  A union bound over at most $n$ positions shows that the two orders differ with probability at most $n/N^2$.  Since the weighted load lies in $[0,n]$, the expected weighted loads differ by at most $n^2/N^2\le 1$.  This gives the claimed load bound.

Each point of $P$ is drawn with probability~$1/N$, so \eqref{eq:finite-estimate} has conditional mean~$L_{\pi,A}(p)/(N+1)$ given $(\pi,A)$.  The estimates lie in $[0,1]$, so \eqref{eq:mwu-potential} applies to the finite sampler.  Writing $g=1+\ln(n/s)$, the argument of Lemma~\ref{lem:fixed-guess} bounds the expected cost by
\[
  (s+1)+s\left(\frac{2(N+1)}{T}\ln \frac ns+2(2+2\kappa g)\right)
  \le (6+4\kappa)s g,
\]
where we used $N+1\le T$ and $s\ge 1$.
The weights still have $O(n\log n)$ bits.  All prefix draws, the index~$J$, and the point or dummy position~$Q$ have sample spaces of power-of-two size.  The remaining draws have a fixed attempt limit.  Thus the number of random bits and the running time apart from edge tests are bounded by polynomials.  There are $T=2N=O(n)$ rounds per guess, so the edge-test count remains $O(n^2\log n)$ over all guesses.  The same budgets and phase analysis therefore give the cost bound in Theorem~\ref{thm:sparse}.
\end{proof}

\section{Finding the first point in a disk}\label{sec:disk-order-queries}

We adapt the binary-tree nearest-neighbor search of Liu and Wang~\cite[Lemma~8]{liuwang} by using the sampled order in place of their order by $x$-coordinate.  For each order prepared by Algorithm~\ref{alg:online}, build a balanced binary tree whose leaves are the points of $P$ in that order.  At each internal node, build the Voronoi diagram of the points in its subtree and preprocess it for point location~\cite{fortune,kirkpatrick}.  For $m$ sites the construction uses $O(m\log m)$ geometric operations and a nearest-neighbor query takes $O(\log m)$ time.  A leaf answers the query using its single site.

The root's point set meets the disk.  At each internal node, find the point in the left subtree nearest to the disk center.  Descend left if this point lies in the disk, and right otherwise.  The current subtree therefore always meets the disk.  Since every point in the left subtree precedes every point in the right subtree, the leaf reached is the first point in the disk.  The search uses $O(\log n)$ nearest-neighbor queries.

The diagrams at each tree level contain at most $n$ sites in total.  Their construction uses $O(n\log^2 n)$ geometric operations per order, or $O(n\log^3 n)$ over all orders, within the $O(n^2\log n)$ preprocessing bound.  Each request makes at most two searches, one in the current order and one in the next order if the phase changes.  Record the last phase in which each point was counted to support constant-time membership tests in $X_s$, and store $X$ in a Boolean array.  Thus each request takes $O(\log^2 n)$ worst-case time.

\section{Lower bound for balls in graphs of maximum degree three}\label{sec:degree3}

Given an online hitting-set instance, we build a graph with unit edge lengths and one candidate vertex for each element of the ground set.  Each range in the original family is the trace of a ball on these vertices, and all the balls have the same radius.

One can also obtain such a graph from a sufficiently large cubic expander, whose clique minor~\cite[Theorem~8.4]{krivelevich-expanders} contains a model of the bipartite incidence graph.  Use unit lengths in rooted spanning trees of the branch sets and a common integer length greater than twice the maximum tree height for incidence edges.  Set the radius to the incidence length plus twice the maximum tree height, so a member is reachable across one incidence edge and a nonmember requires at least three such edges within the model.  Give unused edges integer lengths greater than the radius and subdivide all edges according to their lengths.  This realizes the same traces in a unit-edge graph of maximum degree three.  We give a direct, self-contained construction below.

Start with the bipartite incidence graph joining each element to the ranges that contain it.  Replace each nonisolated vertex by a rooted binary tree with a distinct leaf for each incident edge, and leave isolated vertices unchanged.  Let $h\ge 1$ be an integer upper bound on the tree heights.  For each edge of the incidence graph, join the corresponding leaves by a path of length~$L=2h+1$ with new internal vertices.  These paths attach at distinct leaves, so the resulting graph has maximum degree at most three.  Use the roots of the element trees as candidate vertices.

Represent a range by the ball of radius~$R=L+2h$ centered at its tree's root.  For an element in the range, the connecting path and at most $h$ edges in each tree give a path between the roots of length at most $R$.  For an element outside the range but in the same component, any path between the roots alternates between element trees and range trees.  Since the element does not belong to the range, such a path uses at least three connecting paths and has length at least $3L>R$.

To make the host graph connected, join one vertex of each component to a distinct leaf of a new binary tree by a path of length~$R+1$.  In a nonsingleton component, attach the new path to an internal vertex of a length-$L$ path.  In a singleton component, attach it to its only vertex.  The attachment vertices have degree at most two, so the maximum degree remains at most three.  Distances within a component are unchanged: any walk that leaves and returns to it passes through the same attachment vertex.  Distinct original components remain at distance greater than $R$.

An online hitting-set algorithm for these balls selects candidate roots, which correspond to elements of the original instance.  Presenting the ball corresponding to each arriving range to this algorithm preserves the cost, feasibility, and offline optimum.  Thus the $\Omega(\sqrt n)$ lower bound on the competitive ratio of randomized algorithms against an oblivious adversary holds for these ball instances~\cite{doron}.  The number of vertices in the host graph can be exponential in $|P|$, since it contains a tree for each range.

\section{Refined bounds for online interval hitting}\label{app:refined-intervals}

We consider nonempty interval requests on an ordered set of $n$ candidates against an oblivious adversary.  For real $x\ge 1$, put $h=\lfloor\log_2 x\rfloor$ and define
\[
  F(x)=1+\frac h2+\frac{x-2^h}{2^{h+1}}.
\]
Thus $F(2^h)=1+h/2$, and $F$ is linear between consecutive powers of two; moreover, $F(2x)=F(x)+1/2$.

For $k=1$, the construction in Theorem~\ref{thm:interval-lower} uses $2^h$ candidates and gives a lower bound of $1+h/2$.  For the refined lower bound, we adapt the fair-halving argument of Br\^anzei and Peres~\cite[Theorem~6]{branzei-peres} for prediction with one perfect expert.  We start with all $n$ candidates and choose uniformly between two nearly equal halves at each split.  The probability of needing one more split after depth~$h$ is $(n-2^h)/2^h$, giving the extra term $(n-2^h)/2^{h+1}$ in $F(n)$.  We give a matching upper bound for $k=1$.  Section~\ref{app:refined-intervals-2} extends the refined lower bound to arbitrary $k$.

\subsection{Exact randomized bound for a common candidate}\label{app:refined-intervals-1}

The algorithm goes as follows.  Identify each candidate with its index in the given order, so requests are consecutive blocks in $\{1,\ldots,n\}$.  Draw one shift~$s$ uniformly from $\{0,\ldots,2^{\lceil\log_2 n\rceil}-1\}$.  We assign a rank to each candidate~$i$ as follows.  Start with $i+s$ and count how many times it can be divided by two before reaching an odd integer.  For example, a shifted index of $12=2^2\cdot 3$ has rank~$2$.

Order the candidates by decreasing rank, breaking ties by their original order.  Maintain the running intersection~$J$ of the requests seen so far.  Since the requests share a candidate, $J$ is always nonempty.  After each request, add the first point of $J$ in this order to the selected set, keeping all earlier selections.  Figure~\ref{fig:shifted-rank-example} illustrates the selections for one shift.

\medskip
\noindent\begin{minipage}{\linewidth}
  \captionsetup{type=figure}
  \centering
  \begin{tikzpicture}[x=.95cm,y=.91cm,font=\small,text=overviewink]
    \node[anchor=east] at (.58,0) {Candidate};
    \node[anchor=east] at (.58,-.48) {Rank};
    \foreach \candidate/\rankValue in {1/1,2/0,3/2,4/0,5/1,6/0,7/3,8/0} {
      \node at (\candidate,0) {$\candidate$};
      \node[text=overviewslate] at (\candidate,-.48) {$\rankValue$};
      \draw[overviewslate!24,densely dotted] (\candidate,-.77) -- (\candidate,-3.65);
    }
    \foreach \step/\reqLeft/\reqRight/\intLeft/\intRight/\winner/\action in {
      1/1/6/1/6/3/{Add $3$},
      2/2/8/2/6/3/{No new point},
      3/4/7/4/6/5/{Add $5$}
    } {
      \pgfmathsetmacro{\requestY}{-1.13-.98*(\step-1)}
      \pgfmathsetmacro{\intersectionY}{\requestY-.36}
      \node[anchor=east,text=overviewslate] at (.58,\requestY) {$R_{\step}$};
      \node[anchor=east,text=overviewblue] at (.58,\intersectionY) {$J_{\step}$};
      \draw[overviewslate,line width=.62pt]
        (\reqLeft,\requestY) -- (\reqRight,\requestY)
        (\reqLeft,{\requestY-.07}) -- (\reqLeft,{\requestY+.07})
        (\reqRight,{\requestY-.07}) -- (\reqRight,{\requestY+.07});
      \draw[overviewblue,line width=.92pt]
        (\intLeft,\intersectionY) -- (\intRight,\intersectionY)
        (\intLeft,{\intersectionY-.07}) -- (\intLeft,{\intersectionY+.07})
        (\intRight,{\intersectionY-.07}) -- (\intRight,{\intersectionY+.07});
      \fill[overvieworange] (\winner,\intersectionY) circle[radius=2pt];
      \node[anchor=west,text=overvieworange] at (8.52,\intersectionY) {\action};
    }
  \end{tikzpicture}
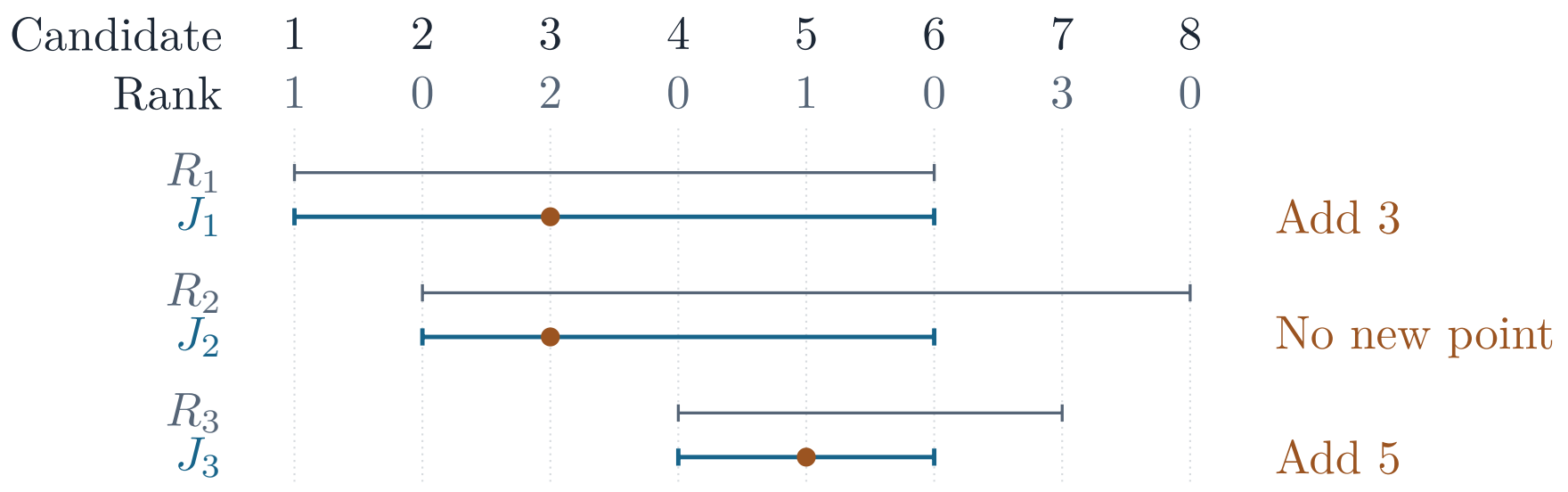
  \caption{The shifted-rank rule for eight candidates and shift~$s=1$.  Gray bars show the requests~$R_j$, and blue bars show the running intersections~$J_j$.  The orange dot marks the maximum-rank candidate at each step.  Point~$3$ remains in the hitting set when the intersection shrinks to~$J_3$.  The final selected set is $\{3,5\}$.}
  \label{fig:shifted-rank-example}
\end{minipage}
\par\medskip

\begin{theorem}\label{thm:interval-common}
The optimal randomized worst-case expected cost for interval sequences with offline optimum one is $F(n)$.
\end{theorem}

\begin{proof}
Every nonempty interval has a unique maximum-rank point.  If two candidates have rank~$r$, their shifted indices are odd multiples of $2^r$.  An even multiple of $2^r$ lies between them, so the corresponding candidate lies in the interval and has higher rank.

An endpoint of an $m$-point interval is its maximum-rank point exactly when the shifted index of the endpoint is divisible by $2^{\lceil\log_2 m\rceil}$.  Indeed, if its rank~$r$ is smaller than $\lceil\log_2 m\rceil$, the point at distance~$2^r$ inward lies in the interval and has higher rank.  If its rank is at least $\lceil\log_2 m\rceil$, the spacing between points with rank that high is at least $m$.

For any fixed nested chain of nonempty intervals inside an $m$-point block, refine the chain from the full block to its final interval of size~$\ell$ by single endpoint deletions.  Choose the deletions from the intervals alone.  Each winner change along the original chain requires a winner change in the refinement.  A deletion from size~$a$ changes the winner with probability $2^{-\lceil\log_2 a\rceil}$, the probability that the deleted endpoint was the winner.  Counting one initial selection, the expected number of selections along the chain is at most
\begin{equation}\label{eq:interval-pivot-cost}
  1+\sum_{a=\ell+1}^m 2^{-\lceil\log_2 a\rceil}
  =1+F(m)-F(\ell).
\end{equation}
Here the equality follows from the slopes of $F$ on consecutive dyadic intervals.  Applying \eqref{eq:interval-pivot-cost} with $m=n$ and $F(\ell)\ge 1$ gives the upper bound.

For the lower bound, start by requesting the full block.  Whenever the current block has size~$m\ge 2$, split it into consecutive halves of sizes $\lfloor m/2\rfloor$ and $\lceil m/2\rceil$ and request either half with probability $1/2$, using independent fair coins.  Stop at a singleton.  The choices can be made before any requests arrive.  Just before a split, the algorithm has at least one selected point in the current block.  Conditional on the history, each such point is discarded with probability $1/2$.  Summing over the splits gives at least half the expected number of splits in expected discarded selections.  Discarded selections are counted once, and the selected point in the terminal singleton contributes one more unit.

Set $h=\lfloor\log_2 n\rfloor$.  At depth~$h$, the $2^h$ possible blocks partition the candidates, and their sizes differ by at most one.  Exactly $n-2^h$ blocks have size two.  Each depth-$h$ block occurs with probability~$2^{-h}$, and the size-two blocks alone require another split.  Thus the expected cost is at least
\[
  1+\frac12\left(h+\frac{n-2^h}{2^h}\right)=F(n).
\]
Every sequence in the distribution has offline optimum one.  Averaging over the sequences gives a fixed sequence on which the expectation over the algorithm's random choices is at least $F(n)$.
\end{proof}

The deterministic competitive ratio is $1+\lfloor\log_2 n\rfloor$.  Its lower bound already holds for nested requests with a common candidate~\cite[Proposition~6]{even}.
For $n$ candidates with offline optimum one, Theorem~\ref{thm:interval-common} gives the randomized cost $F(n)$.  The gap from the deterministic cost is
\[
  (1+h)-F(n)=\frac h2-\frac{n-2^h}{2^{h+1}}.
\]

\subsection{Bounds for arbitrary offline optimum}\label{app:refined-intervals-2}

Proposition~\ref{prop:interval-refined-lower} strengthens Theorem~\ref{thm:interval-lower} when $n/k$ lies strictly between consecutive powers of two.  The bound follows by applying the fair-split construction from the proof of Theorem~\ref{thm:interval-common} independently in $k$ disjoint blocks.

\begin{proposition}\label{prop:interval-refined-lower}
For all integers~$n,k$ with $1\le k\le n$, every randomized online interval-hitting algorithm on $n$ candidates can be forced to select at least $kF(n/k)$ points in expectation by a fixed request sequence with offline optimum~$k$.
\end{proposition}
\begin{proof}
Partition the candidates into $k$ consecutive blocks~$B_1,\ldots,B_k$ whose sizes differ by at most one.  Use the fair-split distribution independently in each block and reveal the resulting chains in block order.  Every sequence has offline optimum~$k$.  The lower-bound argument counts selections separately in the disjoint blocks, including any earlier selections present before a split.  The expected cost over the request distribution and the algorithm's random choices is at least $\sum_{j=1}^k F(|B_j|)$.  The block sizes are $\lfloor n/k\rfloor$ or $\lceil n/k\rceil$ and have mean $n/k$, so linearity of $F$ between consecutive integers gives $kF(n/k)$ for this sum.  Averaging gives a fixed request sequence with expected cost at least this large.
\end{proof}

The realizations in Section~\ref{sec:lower} transfer the bound to disks on collinear candidates and to shortest-path balls on paths.

The upper bound runs the shifted-rank rule of Theorem~\ref{thm:interval-common} in contiguous regions, each with its own running intersection.  To limit the cost as an intersection shrinks, the algorithm splits its region and restarts the rule in the children.  The points at which it splits are called \emph{separators}.  It adds them to the hitting set~$X$ and records them in the set~$Z$ of separators.  The maximal consecutive blocks left after removing the separators become the child regions.

Fix an integer~$b\ge 1$.  Start with one region~$P$ and sample a single shift for all regions.  Each region~$B$ starts with running intersection $J_B=B$ and an empty request history.  A \emph{phase} consists of the requests processed in a region through its split or the end of the sequence.

A request containing a separator is already hit.  Every other request~$R$ lies in one current region~$B$.  Compute the tentative intersection $I=J_B\cap R$ even when a previous shifted-rank selection already hits $R$.  Continue the shifted-rank rule in $B$ if $|I|>|B|/2^b$.  Otherwise, split $B$ as illustrated in Figure~\ref{fig:interval-splits} and specified in Algorithm~\ref{alg:interval-splitting}.
\begin{itemize}
\item If $I\ne\varnothing$, its endpoints hit the triggering request.  If a middle child~$B'$ exists, every request processed in the parent contains all of~$B'$.  Thus the usual initialization $J_{B'}=B'$ already gives the common intersection of those requests within the child.
\item If $I=\varnothing$, the triggering request lies wholly to one side of $J_B$.  Among previous requests in $B$ that attain the endpoint of $J_B$ on that side, choose the earliest as the disjoint witness.
\end{itemize}

\medskip
\noindent\begin{minipage}{\linewidth}
  \captionsetup{type=figure}
  \centering
  \begin{subfigure}[t]{.47\linewidth}
    \centering
    \begin{tikzpicture}[x=.40cm,y=.73cm,font=\small,text=overviewink]
      \useasboundingbox (.4,-.75) rectangle (14.6,2.57);
      \draw[overviewslate!65,line width=.5pt] (.66,0) -- (14.34,0);
      \foreach \cut in {5,10} {
        \draw[overvieworange!70,densely dashed] (\cut,-.21) -- (\cut,1.02);
      }
      \foreach \candidate in {1,...,14} {
        \draw[overviewink, fill=white] (\candidate,0) circle[radius=1.35pt];
      }
      \foreach \cut in {5,10} {
        \fill[overvieworange] (\cut,0) circle[radius=2.15pt];
      }
      \foreach \reqLeft/\reqRight/\reqY/\reqLabel in {2/10/2/R_1,5/13/1.21/R_2} {
        \draw[overviewslate,line width=.65pt]
          (\reqLeft,\reqY) -- node[midway,above=1pt] {$\reqLabel$} (\reqRight,\reqY)
          (\reqLeft,{\reqY-.075}) -- (\reqLeft,{\reqY+.075})
          (\reqRight,{\reqY-.075}) -- (\reqRight,{\reqY+.075});
      }
      \draw[overviewblue,line width=.9pt, |-|]
        (5,.52) -- node[midway,above=1pt] {$I$} (10,.52);
      \node[text=overviewslate] at (2.5,-.47) {Left};
      \node[text=overviewblue] at (7.5,-.47) {Middle};
      \node[text=overviewslate] at (12.5,-.47) {Right};
    \end{tikzpicture}
    \caption{A small nonempty intersection.}
    \label{fig:interval-split-small}
  \end{subfigure}\hfill
  \begin{subfigure}[t]{.47\linewidth}
    \centering
    \begin{tikzpicture}[x=.40cm,y=.73cm,font=\small,text=overviewink]
      \useasboundingbox (.4,-.75) rectangle (14.6,2.57);
      \draw[overviewslate!65,line width=.5pt] (.66,0) -- (14.34,0);
      \draw[overvieworange!70,densely dashed] (8,-.21) -- (8,1.91);
      \foreach \candidate in {1,...,14} {
        \draw[overviewink, fill=white] (\candidate,0) circle[radius=1.35pt];
      }
      \fill[overvieworange] (8,0) circle[radius=2.15pt];
      \draw[overviewslate,line width=.65pt, |-|]
        (1,2) -- node[midway,above=1pt] {$W_L$} (8,2);
        \draw[overviewslate,line width=.65pt, |-|]
        (10,2) -- node[midway,above=1pt] {$R$} (14,2);
      \draw[overviewblue,-{Latex[length=1.55mm]},line width=.7pt]
        (8.85,.55) -- node[midway,above=1pt] {Forward $R$} (14,.55);
      \node[text=overviewslate] at (4,-.47) {Left};
      \node[text=overviewblue] at (11.5,-.47) {Right};
    \end{tikzpicture}
    \caption{An empty intersection.}
    \label{fig:interval-split-empty}
  \end{subfigure}
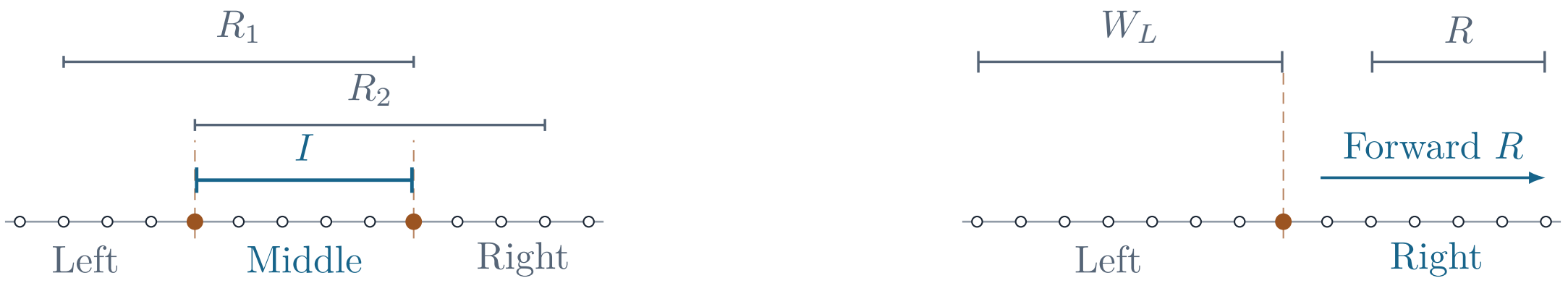
  \caption{The split cases for a region of $m=14$ candidates and $b=1$.  Orange dots mark separators.  In (a), $I=R_1\cap R_2$.  In (b), the earlier request~$W_L$ is the left witness and the arriving request~$R$ is the right witness, so $R$ is forwarded to the right child.}
  \label{fig:interval-splits}
\end{minipage}
\par\medskip

\begin{pseudocode}
  \pseudocodetitle{alg:interval-splitting}{Online interval hitting by region splitting}
  \begin{tabbing}
    0.\quad\=\quad\=\quad\=\quad\=\kill
    1.\> Sample one shift.  Set $X=Z=\varnothing$ and initialize the sole region $P$.\\
    2.\> \textbf{for} each arriving interval $R$ with $R\cap Z=\varnothing$ \textbf{do}\\
    \> Let $B$ be the current region containing $R$.\\
    \> \textbf{repeat}\\
    \>\> $I\gets J_B\cap R$\\
    \>\> \textbf{if} $|I|>|B|/2^b$ \textbf{then}\\
    \>\>\> Set $J_B\gets I$ and record $R$ in the request history of $B$.\\
    \>\>\> Add the maximum-rank point of $I$ to $X$ and \textbf{break}.\\
    \>\> \textbf{if} $I\ne\varnothing$ \textbf{then}\\
    \>\>\> Split $B$ at the endpoints of $I$ and \textbf{break}.\\
    \>\> Use an endpoint of $J_B$ to find the earlier witness disjoint from $R$.\\
    \>\> Let $z$ be the right endpoint of the left witness and split $B$ at $z$.\\
    \>\> \textbf{if} $R$ is the left witness \textbf{then break}.\\
    \>\> Set $B$ to the right child of the split.\\
    3.\> \textbf{return} $X$.
  \end{tabbing}
\end{pseudocode}

A forwarded request enters a fresh region, so its next tentative intersection equals the request itself.  One of the nonempty cases then hits it.

\begin{theorem}\label{thm:interval-splitting}
For each integer~$b\ge 1$, there is a randomized online interval-hitting algorithm on $n$ candidates.  On every fixed nonempty request sequence with offline optimum~$k$, its expected cost is at most
\[
  \left(\frac b2+3\right)\left(\frac{k\log_2(n/k)}{b}+2k-1\right).
\]
\end{theorem}
\begin{proof}
Fix a request sequence.  The algorithm skips an arriving request exactly when it contains a separator, and the requests processed in a region determine when and where it splits.  Thus the regions, their intersection chains, and the separators are independent of the shift.  We can therefore apply \eqref{eq:interval-pivot-cost} to the shifted-rank selections in each phase.

If a phase in a region of size~$m$ makes any shifted-rank selections, its last nonsplitting intersection has size~$\ell>m/2^b$.  When $m/2^b\ge 1$, the scaling identity for $F$ gives $F(m)-F(\ell)\le b/2$.  When $m/2^b<1$, use $F(m)\le F(2^b)=1+b/2$ and $F(\ell)\ge 1$.  Thus the shifted-rank rule contributes at most $1+b/2$ distinct selections in expectation.  Adding at most two cut purchases gives an expected cost of at most $b/2+3$ per phase.

Fix an optimal hitting set~$\OPT$ of size~$k$.  Each active phase processes a real request contained in its region, so the region contains a point of $\OPT$.  Treat each separator as a singleton child of the split that creates it.  In an empty-intersection split, the left witness lies in the strict left region together with the separator, and the right witness lies in the strict right region.  Since $\OPT$ hits both witnesses, at least two children contain optimum points.  In a nonempty-intersection split, the strict left exterior misses the request attaining the left endpoint of $I$, and the strict right exterior misses the request attaining the right endpoint.  Thus if all $t$ optimum points of a parent pass to one region child, it is the strict middle, of size at most $m/2^b$.

Give a region of size~$m$ containing $t>0$ optimum points potential $t\log_2(m/t)$, and give regions with no optimum points potential zero.  If the child sizes and optimum counts are $m_i$ and $t_i$, concavity of $\log_2$ gives
\[
  \sum_{i:t_i>0}t_i\log_2(m_i/t_i)
  \le t\log_2\!\left(\frac{\sum_{i:t_i>0}m_i}{t}\right)
  \le t\log_2(m/t).
\]
Singleton children have zero potential.  When all $t$ optimum points stay in the middle, the potential drops by at least $tb\ge b$.  Write $L=\log_2(n/k)$.  The initial potential is $kL$, so there are at most $kL/b$ such phases.

To count the remaining active phases, prune the region tree by retaining the regions and separator singletons that contain at least one point of $\OPT$.  Every retained node has all its ancestors retained.  The leaves are pairwise disjoint and each contains a point of $\OPT$, so there are at most $k$ leaves.  The remaining active phases fall into three cases.

\begin{enumerate}
  \item An unsplit active phase is itself a region leaf.
  \item A splitting phase whose optimum points all fall in one separator singleton has $t=1$ and is the unique parent of that separator leaf.
  \item A splitting phase with at least two retained children is a branching node.  Every empty-intersection split falls into this case.
\end{enumerate}

Cases (1) and (2) charge to distinct leaves, giving at most $k$ phases between them.  A rooted tree with at most $k$ leaves has at most $k-1$ branching nodes, so case (3) contributes at most $k-1$ phases.  Hence there are at most $2k-1$ remaining active phases.

Summing the expected costs of the phases gives
\[
  \E[\mathrm{cost}]
  \le \left(\frac b2+3\right)\left(\frac{kL}{b}+2k-1\right)
  \le \frac{kL}{2}+k\left(b+\frac{3L}{b}+6\right).\qedhere
\]
\end{proof}

Taking $b=\max\{1,\lceil\sqrt{\log_2 n}\rceil\}$ gives a competitive ratio of $\tfrac12\log_2 n+O(\sqrt{\log n})$ without knowledge of $k$.  The lower bound $F(n)$ for $k=1$ gives the same leading constant.  If $k$ is known, taking $b=\lceil\sqrt{1+\log_2(n/k)}\rceil$ gives expected cost $\tfrac12 k\log_2(n/k)+O(k\sqrt{1+\log(n/k)})$.

\end{document}